\documentclass[prd,twocolumn,superscriptaddress,floatfix,nofootinbib]{revtex4-2}
\pdfoutput=1 
\usepackage[T1]{fontenc} 
\usepackage{amsmath,amssymb,amsfonts,amsthm}
\usepackage{graphicx}
\usepackage[usenames,dvipsnames]{color}
\usepackage{enumitem}
\usepackage{verbatim}
\usepackage[percent]{overpic}
\usepackage{rotating}
\usepackage{hyperref}
\usepackage{array}
\usepackage{mathtools}
\usepackage{lmodern}
\usepackage[normalem]{ulem}
\usepackage{braket}
\usepackage{tensor}
\usepackage{dsfont}
\usepackage{bm}
\usepackage{tikz}
\usepackage{bbm}
\usepackage{float}
\usepackage{comment}

\usepackage{mathrsfs}

\usetikzlibrary{decorations.pathmorphing}
\usepackage{CJKutf8}

\newcommand{\tcb}{\textcolor{blue}}

\newcommand{\kgy}[1]{{\color{blue}\bf [Ken: #1]}}

\usetikzlibrary{positioning}
\usetikzlibrary{calc,through,backgrounds}

\theoremstyle{definition}
\newtheorem{definition}{Definition}[section]
\newtheorem{theorem}{Theorem}[section]

\newcommand{\kagikako}[1]{\left[ #1 \right]}
\newcommand{\tsx}[1]{ _{\text{#1}} }
\newcommand{\ts}[1]{_\textsc{#1}}
\newcommand{\sst}[1]{^{\text{#1}}}
\newcommand{\ssc}[1]{^{\textsc{#1}}}

\newcommand{\wext}{\braket{W_{\text{ext}}}}
\newcommand{\resh}[1]{\mathcal{F}_{\textsc{i}}(#1)}
\newcommand{\resc}[1]{\mathcal{F}_{\textsc{ii}}(#1)}
\newcommand{\delph}[1]{\delta p\ssc{i}_{#1}}
\newcommand{\delpc}[1]{\delta p\ssc{ii}_{#1}}
\newcommand{\tempfirst}[1]{T\sst{eff}\ts{i}(#1)}
\newcommand{\tempsec}[1]{T\sst{eff}\ts{ii}(#1)}

\allowdisplaybreaks

\DeclareMathOperator{\Tr}{Tr}
\newcommand{\R}{\mathbb{R}}
\newcommand{\dd}{\text{d}}

\newcommand{\bx}{{\bm{x}}}

\newcommand{\id}{\mathbbm{1}}
\newcommand{\sx}{\mathsf{x}}

\newcommand{\ii}{\mathsf{i}}

\begin{document}

\title{Relativistic quantum Otto heat engine using a three-level Unruh-DeWitt detector}

\author{Tomoya Hirotani}
\email{hirotani.tomoya.937@s.kyushu-u.ac.jp}
\affiliation{Department of Physics, Kyushu University, 
744 Motooka, Nishi-Ku, Fukuoka 819-0395, Japan}

\author{Kensuke Gallock-Yoshimura}
\email{gallockyoshimura@biom.t.u-tokyo.ac.jp} 

\affiliation{
Department of Information and Communication Engineering, 
Graduate School of Information Science and Technology,
The University of Tokyo, 
7–3–1 Hongo, Bunkyo-ku, Tokyo 113–8656, Japan
}
\affiliation{Department of Physics, Kyushu University, 
744 Motooka, Nishi-Ku, Fukuoka 819-0395, Japan}

\begin{abstract}
In this study, we explore a relativistic quantum Otto heat engine with a qutrit as the working substance interacting with a quantum scalar field in curved spacetime. 
Unlike qubits, which extract work by simply expanding or shrinking a single energy gap, qutrits allow multiple energy gaps to be adjusted independently, enabling more versatile work extraction in the quantum Otto cycle. 
We derive a general positive work condition in terms of the effective temperature that each pair of energy levels perceives. 
Moreover, we discuss additional subtleties that are absent when using a qubit, such as the generation of coherence terms in the density matrix due to interactions. 
\end{abstract}

\maketitle
\flushbottom

\section{Introduction}
Quantum thermodynamics is an emerging field at the intersection of quantum mechanics and thermodynamics. 
It aims to extend thermodynamic principles within the framework of quantum theory \cite{Vinjanampathy_2016,Goold_2016,gemmer2009quantum}. 
One of the central topics in quantum thermodynamics is the study of quantum heat engines (QHEs), which are the quantum analogs of classical heat engines. 
For example, the \textit{quantum Otto engines} (QOEs) \cite{Scovil.heat.engine.maser, Feldmann.QOE.2000, Kieu.secondlaw.demon.otto, Kieu.heat.engine.2006, Rostovtsev.Otto.2003, QHE.Quan.qutrit, Quan.QHE.2007}, which consist of two adiabatic and two ishochoric processes, have been studied extensively. 
Typically, these QHEs employ a quantum system (such as a qubit) as a working substance immersed in a heat bath. 
Unlike their classical counterparts, QHEs exploit quantum features to achieve performance that would be unattainable in classical systems. 
For example, entanglement and coherence \cite{Marlan_2003, PhysRevLett.89.180402, Dillenschneider_2009,  Zhang_fourlevel_entangle_2007, Uzdin_2015,ozdemir2020quantum,yin2020work,zhang2008entangled, he2012thermal,hardal2015superradiant} can serve as resources for QHEs. 
Furthermore, squeezing the heat bath may also improve performance \cite{PhysRevLett.112.030602, PhysRevE.86.051105}, potentially exceeding the thermal efficiency of classical heat engines.


Recent studies have opened new avenues for combining relativity and quantum thermodynamics, especially in the context of relativistic quantum information (RQI) and QHEs \cite{UnruhOttoEngine, Finn.UnruhOtto, Gallock2023Otto, Xu.UnruhOtto.degenerate, Kane.entangled.Unruh.Otto, Barman.entangled.UnruhOtto, Mukherjee.UnruhOtto.boundary, Chattopadhyay_2019, PhysRevE.86.061108, Sukamto_2023,Myers_2021, Purwanto_2016, PhysRevE.94.022109, Gallock.Otto.delta, Kollas.Otto.2024}. 
In this context, the working substance is modeled by an Unruh-DeWitt (UDW) particle detector \cite{Unruh1979evaporation, DeWitt1979}, which is a two-level quantum system coupled to a quantum scalar field in (curved) spacetime. 
Thus, in relativistic QHEs, the UDW detector extracts thermodynamic work from the quantum scalar field. 
One of the central themes in this area is the use of the Unruh effect. 
The Unruh effect states that a UDW detector uniformly accelerated in the Minkowski vacuum thermalizes at the Unruh temperature $T\ts{U}=\hbar a/2\pi c k\ts{B}$, where $a$ is the acceleration. 
Hence, one can operate a thermodynamic cycle by adjusting the magnitude of the detector's acceleration \cite{UnruhOttoEngine, Finn.UnruhOtto, Gallock2023Otto}.

Although relativistic QHEs employing a two-level working substance have been extensively studied, relativistic QHEs with a multilevel system remain unexplored. 
In fact, multilevel UDW detectors are less commonly used in RQI. 
While two-level UDW detectors are a natural choice due to their simplicity and relevance to the fundamental aspects of light-matter interaction \cite{PhysRevD.87.064038, Funai.lightmatter.2019, Lopp.lightmatter.2021, Shah.UDW.spin.2025}, the use of multilevel UDW detectors is also well-motivated from the perspective of quantum optics. 
In particular, three-level systems have been extensively studied in the context of cavity QED (see e.g., \cite{You:2011}), thereby serving as a physically meaningful extension of the UDW model for exploring richer light-matter interactions in RQI.
As an example, the Unruh effect for a three-level (i.e., qutrit-type) UDW detector was only recently investigated \cite{Lima.Unruh.qutrit}, which showed that subtle issues arise during the thermalization of multilevel UDW detectors. 
Moreover, a gapless multilevel UDW model was used in the context of acceleration radiation in \cite{Ken.gapless.qudit.2025}. 
The aim of this paper is to examine the properties of three-level UDW detectors in the context of QHEs. 
In particular, we derive the positive work condition (PWC) for the relativistic quantum Otto engine (RQOE) using a three-level UDW detector in a globally hyperbolic curved spacetime.

In the context of (nonrelativistic) quantum thermodynamics, Refs.~\cite{QHE.Quan.qutrit, Quan.QHE.2007, Zhang_fourlevel_entangle_2007, sonkar2024energygap, Anka.multilevel.2021} have examined work extraction using multilevel systems. 
In particular, Ref.~\cite{QHE.Quan.qutrit} showed that, in a specific case, the qutrit-PWC is less restrictive than that of two-level systems. 
In this paper, we derive a more general expression for the qutrit-PWC in terms of the effective temperatures perceived by pairs of energy eigenstates. 
We show that the qutrit-PWC can be either less or more restrictive than the qubit-PWC depending on the choice of parameters. 
Indeed, a qutrit can extract positive work even when one pair of energy eigenstates violates the qubit-PWC.

This paper is organized as follows. 
In Sec.~\ref{sec:RQOE}, we introduce our three-level UDW detector and compute the work and heat exchanged during a cycle. 
We then show our main result, the PWC for qutrit, in Sec.~\ref{sec:results}. 
It is compared to the well-known qubit-type PWC, as well as those for qutrits examined in the previous papers \cite{QHE.Quan.qutrit, sonkar2024energygap}. 
We finally conclude in Sec.~\ref{sec:conclusion} and discuss the subtleties associated with the three-level UDW detectors. 
Throughout this paper, we use natural units $\hbar = c= k\ts{b}=1$ and the mostly-plus metric signature convention $(-,+,+,\ldots, +)$.
A spacetime point is denoted by $\sx$.

\section{Relativistic Quantum Otto engine}\label{sec:RQOE}

\subsection{Three-level UDW detector}
Let us introduce a three-level UDW detector. 
Unlike two-level detectors, three-level UDW detectors (and more generally, any $d$-level detectors) can be modeled in various ways, such as employing the spin-1 representations of $SU(2)$ or the Heisenberg-Weyl model \cite{Lima.Unruh.qutrit}. 
These models differ in the allowed transitions. 
In this paper, we focus on the qutrit model using the spin-1 representations of $SU(2)$.\footnote{In quantum optics, this model corresponds to the $\Xi$-type atom (also known as the ladder or cascade type). }

For the sake of generality, consider a pointlike three-level UDW detector traveling along an arbitrary timelike trajectory in an $(n+1)$-dimensional globally hyperbolic spacetime. 
The total Hamiltonian in the Schr\"{o}dinger picture $\hat H\tsx{S,tot}$ is written as
\begin{equation}
    \hat H\tsx{S,tot} 
    = 
        \hat H\ts{S,d} +\hat H_{\text{S},\phi} +\hat H\tsx{S,int}\,.
\end{equation}
Here, $\hat H\ts{S,d}$ is the free Hamiltonian of the detector given by
\begin{equation}
    \hat H\ts{S,d}
    =
        \epsilon_0 \ket{e_0}\bra{e_0} + \epsilon_1 \ket{e_1}\bra{e_1} + \epsilon_2 \ket{e_2}\bra{e_2} \,,
    \label{Free Hamiltonian of qutrit ep}
\end{equation}
where $\ket{e_0}$ is the ground state, and $\ket{e_1}$ and $\ket{e_2}$ are the first and second excited states, respectively. 
Each eigenstate $\{ \ket{e_0}, \ket{e_1}, \ket{e_2} \}$ has the respective energy eigenvalue denoted by $\epsilon_0$, $\epsilon_1$ and $\epsilon_2$. 
In what follows, we express the detector's free Hamiltonian \eqref{Free Hamiltonian of qutrit ep} using the energy gaps $\Omega_{01} \equiv \epsilon_1 - \epsilon_0$, $\Omega_{02} \equiv \epsilon_2 - \epsilon_0$, and $\Omega_{12} \equiv \epsilon_2 - \epsilon_1$: 
\begin{equation}
    \hat H\ts{S,d}
    =
        \Omega_{01} \ket{e_1}\bra{e_1} 
        + \Omega_{02} \ket{e_2}\bra{e_2}\,,
\end{equation}
where $\epsilon_0=0$ was assumed as there is no loss of generality. 
Note that if we set $\Omega_{01}=\Omega_{02} (\equiv \Omega)$ then the free Hamiltonian corresponds to $\hat H\ts{S,d}=\Omega (\hat J_z + \id)$, where $\hat J_z$ is the $z$-component of the spin-1 angular momentum operator. 
In this case, each of our energy eigenstate $\{ \ket{e_0}, \ket{e_1}, \ket{e_2} \}$ corresponds to the eigenstate of $\hat J_z$: $\{ \ket{-1}, \ket{0}, \ket{1} \}$, where $\ket{m}\equiv \ket{j=1, m}$.

\begin{figure}[t]
    \centering
\includegraphics[width=0.6\linewidth]{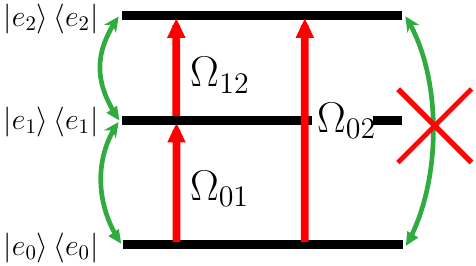}
    \caption{The qutrit transitions represented by the spin-1 operator. }
    \label{fig:spin1 transition}
\end{figure}

$\hat H_{\text{S},\phi}$ is the free Hamiltonian of the quantum scalar field, and the interaction Hamiltonian $\hat H\tsx{S,int}$ is given by


\begin{equation}
   \hat H\tsx{S,int} 
   = 
    \lambda \chi(\tau/\sigma) \hat J_x \otimes \hat \phi (\bx)\,,
\end{equation}
where $\lambda$ is the coupling constant between the qutrit and the field $\hat \phi$, and switching function $\chi(\tau / \sigma)$ determines the time dependence of coupling with $\sigma$ being the typical time scale of interaction. 
This interaction Hamiltonian is a natural extension of the two-level UDW model by replacing the Pauli matrix $\hat \sigma_x$ with the spin-1 angular momentum operator, 
\begin{align}
    \hat J_x 
    &=
        \frac{1}{\sqrt{2}}(\ket{e_1}\bra{e_0}+\ket{e_0}\bra{e_1}+\ket{e_2}\bra{e_1}+\ket{e_1}\bra{e_2})\,,
\end{align}
which determines how the qutrit transitions through the interaction.
As shown in Fig.~\ref{fig:spin1 transition}, this represents transitions to adjacent energy levels, such as between $\ket{e_0} \leftrightarrow \ket{e_1}$ and $\ket{e_1}\leftrightarrow \ket{e_2}$, but not to one skipped level, such as between $\ket{e_0}\leftrightarrow \ket{e_2}$.

In the interaction picture, the interaction Hamiltonian (denoted by $\hat H\ts{I}$) reads 
\begin{align}
    \hat H\ts{I} (\tau)
    =
        \lambda
        \chi (\tau/\sigma) \hat J_x(\tau) \otimes \hat \phi(\sx(\tau))\,,
\end{align}
where 
\begin{align}
    \hat J_x(\tau)
    &\coloneqq
        \dfrac{1}{\sqrt{2}}
        (
            e^{\ii \Omega_{01}\tau} \ket{e_1}\bra{e_0} 
            + 
            e^{-\ii \Omega_{01}\tau} \ket{e_0}\bra{e_1} \notag  \\
            & \qquad
            + 
            e^{\ii \Omega_{12}\tau} \ket{e_2}\bra{e_1} 
            + 
            e^{-\ii \Omega_{12}\tau} \ket{e_1}\bra{e_2}
        )\,.
\end{align}
Here, $\tau$ is the proper time of the UDW detector, and $\hat \phi(\sx(\tau))$ is the field operator defined along the detector's timelike trajectory $\sx(\tau)$. 
The time evolution operator is expressed using the time ordering $\mathcal{T}_{\tau}$ with respect to the proper time $\tau$:
\begin{equation}
    \hat U\ts{I} 
    = 
        \mathcal{T}_{\tau} 
        \exp 
        \left( 
            -\ii \int_{\R} \dd \tau\,
            \hat H\ts{I} (\tau) 
        \right) .
\end{equation}
Assuming the coupling constant is small, we perform the Dyson series expansion so that 
\begin{subequations}
\begin{align}
    \hat U\ts{I} 
    &= 
        \id
        +
        \hat U\ts{I}^{(1)} + \hat U\ts{I}^{(2)} + \mathcal{O}(\lambda^3)\,, \\
    \hat U\ts{I}^{(1)} 
    &= 
        -\ii 
        \int_\R \dd\tau\,
        \hat H\ts{I} (\tau)\,, \label{eq:U1} \\
    \hat U\ts{I}^{(2)} 
    &= 
        - \int_\R \dd\tau 
        \int_\R \dd\tau' \,
        \Theta(t(\tau) - t(\tau'))
        \hat H\ts{I} (\tau) \hat H\ts{I} (\tau')\,, \label{eq:U2}
\end{align}
\end{subequations}
where $\Theta$ denotes the Heaviside step function.


\begin{figure*}[t]
    \centering
    \includegraphics[width=\linewidth]{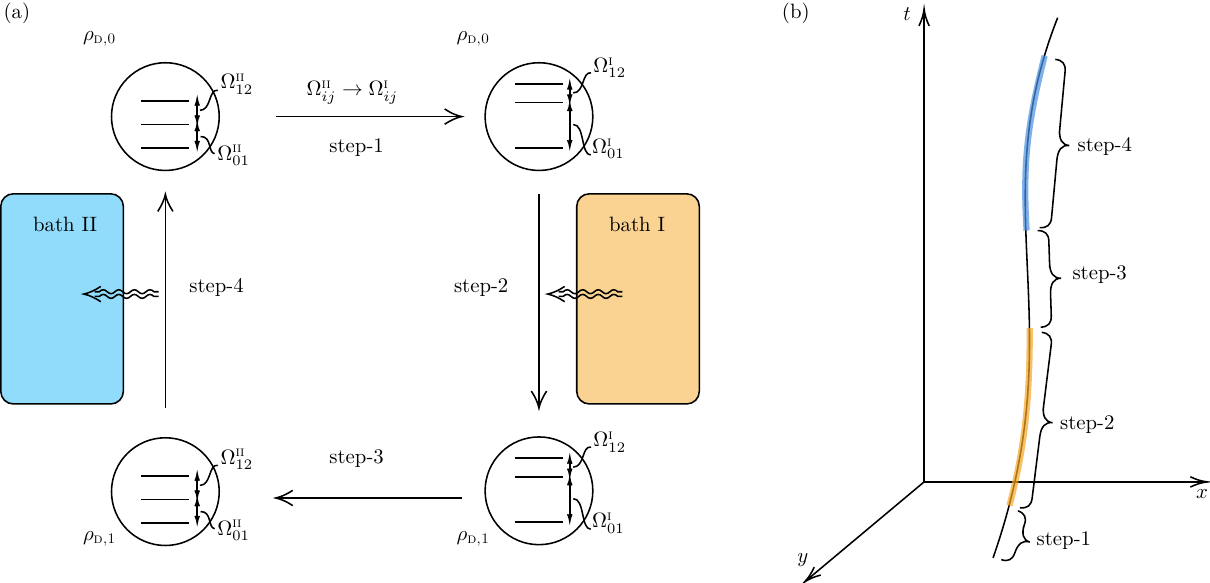}
    \caption{
    (a) A schematic diagram for the quantum Otto cycle using a three-level quantum system. 
    (b) The corresponding relativistic version of the cycle in a spacetime diagram. 
    The vertical curve represents the spacetime trajectory of the three-level UDW detector. 
    }
    \label{fig:RQOE}
\end{figure*}

\subsection{Cycle}
Based on the above setup, we now review the relativistic quantum Otto engine (RQOE) depicted in Fig.~\ref{fig:RQOE} and calculate the work and heat exchanged in each stroke. 
The RQOE consists of two subsystems: a working substance and a bath. In this study, the qutrit serves as the working substance, while the quantum field acts as the bath. The bath that interacts with the qutrit in the first isochoric process (step 2) is referred to as bath I, and the one that interacts with the qutrit in the second isochoric process (step 4) is referred to as bath II.
Below, we assume that the initial state of the joint systems is the product
\begin{equation}
    \rho\tsx{tot,0} = \rho\ts{d,0}\otimes \rho_{\phi,0}\,,
\end{equation}
where $\rho\ts{d,0}$ and $\rho_{\phi,0}$ are the initial states of the detector and the field, respectively. 
Here, the detector is initially prepared in a diagonal state in the basis $\{ \ket{e_2}, \ket{e_1}, \ket{e_0} \}$
\begin{equation}
    \rho\ts{d,0} 
    = 
        \begin{bmatrix}
            p_2 & 0 & 0 \\
            0 & p_1 & 0 \\
            0 & 0 & 1-p_1-p_2 \\
        \end{bmatrix}\,, \label{eq:initial detector state}
\end{equation}
with $p_1, p_2 \in [0,1]$. 
We note that the qutrit's initial state need not be a thermal state. The initial energy gaps are set to $\Omega_{ij}\ssc{ii}$ [see top-left in Fig.~\ref{fig:RQOE}(a)], where II named after `bath II', which the qutrit interacts at the end of the cycle. 
We start with such energy gaps because the cycle must be closed at the end.

The initial state of the field $\rho_{\phi,0}$ is assumed to be a quasifree state, which is a particular class of quantum states whose corresponding $n$-point correlation functions (i.e., Wightman functions) $\braket{\hat \phi(\sx_1) \hat \phi(\sx_2)\ldots \hat \phi(\sx_n)}_{\rho_{\phi,0}}\coloneqq \Tr[ \hat \phi(\sx_1) \hat \phi(\sx_2)\ldots \hat \phi(\sx_n) \rho_{\phi,0} ]$ are expressed by two-point correlation functions only. 
Examples include the vacuum, squeezed, and Kubo-Martin-Schwinger (KMS) thermal states.

\noindent 
(\textit{step-1}) \textit{Quantum adiabatic process}:

The quantum adiabatic process is the process where the energy gaps change very slowly as $\Omega_{ij}\ssc{ii} \to \Omega_{ij}\ssc{i}$ without making a contact with a quantum field. 
We note that $\Omega_{02}\ssc{ii}(=\Omega_{01}\ssc{ii}+\Omega_{12}\ssc{ii}) \to \Omega_{02}\ssc{i}(=\Omega_{01}\ssc{i}+\Omega_{12}\ssc{i})$, and the qutrit's free Hamiltonian during this process is time-dependent:
\begin{equation}
    \hat H\ts{d}(\tau)
    =
        \Omega_{01}\ssc{ii}(\tau) \ket{e_1}\bra{e_1} 
        + \Omega_{02}\ssc{ii}(\tau) \ket{e_2}\bra{e_2}\,.
\end{equation}
Moreover, each energy gap can be either enlarged or shrunk, and we will consider each energy gap configuration later.

If the Hamiltonian is changed slowly enough, the quantum adiabatic theorem \cite{Griffiths2005QM.2nd} ensures that this process can be performed without changing the detector's state $\rho\ts{d,0}$.
Thus, the work  done on the qutrit, $\braket{W_1}$, and the heat it receives, $\braket{Q_1}$, during this process are given by 
\begin{align}
        \langle W_1 \rangle
        &=
            \int \dd \tau 
            \Tr \left[ 
                \rho\ts{d,0} 
                \frac{\dd \hat H\ts{d}(\tau)}{\dd \tau}
            \right] \notag  \\
        &= 
            (p_1+p_2) \Delta \Omega_{01} +p_2 \Delta \Omega_{12}\,, \label{eq:W1} \\
        \langle Q_1 \rangle
        &=
            0\,, 
\end{align}
where $\Delta\Omega_{01} \equiv \Omega_{01}\ssc{i} - \Omega_{01}\ssc{ii}, \Delta\Omega_{12}\equiv \Omega_{12}\ssc{i}-\Omega_{12}\ssc{ii}$ are the change in each energy gap.

\noindent 
(\textit{step-2}) \textit{Quantum isochoric process}:

During this process, the qutrit interacts with a quantum field [bath I in Fig.~\ref{fig:RQOE}(a)] while its energy gaps remain fixed at $\Omega_{ij}\ssc{i}$.  
Note that the detector's free Hamiltonian reads $\hat H\ts{d}=\Omega_{01}\ssc{i} \ket{e_1}\bra{e_1} +\Omega_{12}\ssc{i} \ket{e_2}\bra{e_2}$ throughout this process, thus no work is done: $\braket{W_2}=0$. 
On the other hand, the interaction alters the qutrit's quantum state as $\rho\ts{d,0} \to \rho\ts{d,1}$, which leads to the exchange of heat defined by 
\begin{align}
    \braket{Q_2}
    &\coloneqq
        \int \dd \tau\, 
        \Tr 
        \kagikako{
            \hat H\ts{d} 
            \frac{\dd \rho\ts{d}(\tau)}{\dd \tau}
        } \notag \\
    &=
        \Tr[ \rho\ts{d,1} \hat H\ts{d} ]
        -
        \Tr[ \rho\ts{d,0} \hat H\ts{d} ]\,, \label{eq:Q2 RQOE def}
\end{align}

Let us obtain the post-interaction density matrix $\rho\ts{d,1}$ of the three-level UDW detector. 
We first note that the time dependence of the interaction is specified by a compactly supported switching function $\chi\ts{i}(\tau)$. 
In particular, we will assume that the switching function is symmetric in $\tau$ for simplicity (e.g., rectangular function), and we write $\chi\ts{i}[(\tau-\tau\ts{i})/\sigma\ts{i}]$, where $\sigma\ts{i}$ is the interaction duration and $\tau\ts{i}$ is its midpoint. 
Thus, its support reads $\text{supp}(\chi\ts{i})=[\tau\ts{i}-\sigma\ts{i}/2, \tau\ts{i}+\sigma\ts{i}/2]$.

The density matrix $\rho\ts{d,1}$ is obtained by evolving the total system over time and tracing out the field degrees of freedom:
    \begin{align}
        \rho\ts{d,1}
        &=
            \Tr_{\phi}
            [ \hat U\ts{I}
            \rho\tsx{tot,0} 
            \hat U\ts{I}^{\dag} ]
            \notag \\
        &=
            \rho\ts{d,0}
            + 
            \Tr_{\phi}
            [
                \hat U\ts{I}^{(1)}
                \rho\tsx{tot,0} 
                \hat U\ts{I}^{(1)\dag} 
            ]
            + 
            \Tr_{\phi}
            [
                \hat U\ts{I}^{(2)}
                \rho\tsx{tot,0} 
            ] \notag \\
            &\quad
            +
            \Tr_{\phi}
            [
                \rho\tsx{tot,0} 
                \hat U\ts{I}^{(2)\dag} 
            ]
            + \mathcal{O}(\lambda^4)\,,
    \end{align}
where $\hat U\ts{I}^{(1)}$ and $\hat U\ts{I}^{(2)}$ are given in Eqs.~\eqref{eq:U1} and \eqref{eq:U2}, respectively. 
Since we are assuming that the field's initial state $\rho_{\phi,0}$ is quasifree, only the even-point correlation functions survive in $\rho\ts{d,1}$, leading to its dependence on the even-power in the coupling constant $\lambda$. 

\begin{widetext}
Let us define 
\begin{subequations}
\begin{align}
    \resh{\pm \Omega\ssc{i}_{ij}}
    &\coloneqq
        \frac{1}{\sigma\ts{i}}
        \int_{\R} \dd\tau
        \int_{\R} \dd\tau'\,
        \chi\left( \frac{\tau - \tau\ts{i}}{\sigma\ts{i}}\right)
        \chi\left(\frac{\tau' - \tau\ts{i}}{\sigma\ts{i}}\right) 
        e^{\mp \ii\Omega\ssc{i}_{ij} (\tau-\tau')} W(\sx(\tau), \sx(\tau'))\,, \label{eq:response hot} \\
    \mathcal X\ts{i}
    &\coloneqq
        \frac{1}{\sigma\ts{i}}
        \int_{\R} \dd\tau
        \int_{\R} \dd\tau'\,
        \chi\left( \frac{\tau - \tau\ts{i}}{\sigma\ts{i}}\right)
        \chi\left(\frac{\tau' - \tau\ts{i}}{\sigma\ts{i}}\right) 
        e^{- \ii(\Omega_{12}\ssc{i}\tau + \Omega_{01}\ssc{i} \tau')} 
        W(\sx(\tau), \sx(\tau')) \,, \\
    \mathcal Y\ts{i}
    &\coloneqq
        \frac{1}{\sigma\ts{i}}
        \int_{\R} \dd \tau
        \int_{\R} \dd\tau'\,
        \Theta(t(\tau) - t(\tau'))
        \chi\left( \frac{\tau - \tau\ts{i}}{\sigma\ts{i}}\right)
        \chi\left(\frac{\tau' - \tau\ts{i}}{\sigma\ts{i}}\right) 
        e^{-\ii (\Omega_{01}\ssc{i} \tau + \Omega_{12}\ssc{i} \tau')} 
        W(\sx(\tau), \sx(\tau'))\,, \\
    \mathcal Z\ts{i}
    &\coloneqq
        \frac{1}{\sigma\ts{i}}
        \int_{\R} \dd \tau
        \int_{\R} \dd\tau'\,
        \Theta(t(\tau') - t(\tau))
        \chi\left( \frac{\tau - \tau\ts{i}}{\sigma\ts{i}}\right)
        \chi\left(\frac{\tau' - \tau\ts{i}}{\sigma\ts{i}}\right) 
        e^{\ii (\Omega_{01}\ssc{i} \tau +\Omega_{12}\ssc{i} \tau')} 
        W(\sx(\tau), \sx(\tau'))\,,
\end{align}
\end{subequations}
where $\Omega\ssc{i}_{ij} \in \{ \Omega_{01}\ssc{i}, \Omega_{12}\ssc{i} \}$, $W(\sx(\tau), \sx(\tau'))$ is the pullback of the two-point Wightman function along the detector's trajectory, and $\resh{\pm \Omega\ssc{i}_{ij}}$ is the response function. 
The response function determines the probability of transition between two energy levels of the qutrit. 
For example, $\resh{\Omega_{12}\ssc{i}}$ reflects the excitation probability from $\ket{e_1}$ to $\ket{e_2}$ when their energy gap is $\Omega_{12}\ssc{i}$, while $\resh{-\Omega_{12}\ssc{i}}$ is for the deexcitation $\ket{e_2} \to \ket{e_1}$. 
Similarly, $\resh{\pm \Omega_{01}\ssc{i}}$ is the probability of transition between $\ket{e_0}\leftrightarrow \ket{e_1}$.

The density matrix for the qutrit after the interaction is then \cite{Lima.Unruh.qutrit}
\begin{align}
    \rho\ts{d,1}
    &=
        \begin{bmatrix}
            p_2 + \delph{2} & 0 & \mathcal C\ts{i} \\
            0 & p_1 + \delph{1} & 0 \\
            \mathcal C\ts{i}^* & 0 & 1-p_1-p_2-\delph{1} - \delph{2}
        \end{bmatrix}
        + \mathcal{O}(\lambda^4)\,,
\end{align}
where 
\begin{subequations}
\begin{align}
    \delph{2} 
    &\coloneqq
        \frac{\lambda^2 \sigma\ts{i} }{2}
        [
            p_1 \resh{\Omega_{12}\ssc{i}}
            -
            p_2 \resh{-\Omega_{12}\ssc{i}}
        ]\,, \label{eq:delta p2H} \\
    \delph{1} 
    &\coloneqq
        \frac{\lambda^2 \sigma\ts{i} }{2}
        [
            (1-p_1-p_2) \resh{\Omega_{01}\ssc{i}} 
            - 
            p_1 \resh{-\Omega_{01}\ssc{i}} 
            + 
            p_2 \resh{-\Omega_{12}\ssc{i}} 
            -
            p_1 \resh{\Omega_{12}\ssc{i}} 
        ]\,, \label{eq:delta p1H}\\
    \mathcal{C}\ts{i}
    &\coloneqq
        \frac{\lambda^2 \sigma\ts{i} }{2}
        [
            p_1 \mathcal X\ts{i}
            -
            (1-p_1-p_2) \mathcal{Y}\ts{i}
            - 
            p_2 \mathcal{Z}\ts{i}
        ]\,. \label{eq:hot coherence term}
\end{align}
\end{subequations}

In the case of qutrit-type UDW detectors, an off-diagonal term $\mathcal{C}\ts{i}$ appears, though it is absent in two-level UDW detector models within perturbation theory. 
It was shown in \cite{Lima.Unruh.qutrit} that the off-diagonal coherence term $\mathcal{C}\ts{i}$ vanishes in the adiabatic long-interaction limit if a qutrit is initially prepared in a diagonal state \eqref{eq:initial detector state}.\footnote{While this statement holds for smooth switching functions such as Gaussian switchings, the term $\mathcal{C}\ts{i}$ can be divergent if one employs sudden switchings, even in the long-interaction limit. } 
However, this term is nonvanishing as long as the interaction duration is finite, and it leads to some subtleties in the relativistic quantum Otto engine, which we will address later. 
Nevertheless, since only the diagonal terms in $\rho\ts{d,1}$ contribute to the heat in \eqref{eq:Q2 RQOE def}, $\braket{Q_2}$ is independent of the off-diagonal coherence term: 
\begin{equation}
    \braket{Q_2} 
    =
        (\delph{2} +\delph{1}) \Omega_{01}\ssc{i} 
        + 
        \delph{2} \Omega_{12}\ssc{i} \,.
\end{equation}

\noindent 
(\textit{step-3}) \textit{Quantum adiabatic process}:

Detach the qutrit from the field and restore the energy gaps to its original width: $\Omega_{ij}\ssc{i}\to \Omega_{ij}\ssc{ii}$. 
The heat and work during this process result in 
\begin{align}
    \braket{Q_3}
    &=
        0\,, \\
    \braket{W_3}
    &=
        \int \dd \tau\,
        \Tr 
        \kagikako{
            \rho\ts{d,1} \frac{\dd \hat H\ts{d}(\tau)}{\dd \tau}
        } 
    =
        -(p_1 +p_2 + \delph{1} + \delph{2}) \Delta\Omega\ts{01} 
        -
        (p_2+\delph{2}) \Delta \Omega_{12}\,.
\end{align}

\noindent 
(\textit{step-4}) \textit{Quantum isochoric process}: 

Another isochoric process (i.e., the qutrit interacts with the quantum field) is implemented using $\chi\ts{ii}
[(\tau-\tau\ts{ii})/\sigma\ts{ii}]$ for the switching function. 
To ensure that the support of the interaction region does not overlap with the previous isochoric process, we ensure that the supports of the two switching functions are disjoint by implementing $\tau\ts{i} + \sigma\ts{i}/2 < \tau \ts{ii} + \sigma \ts{ii}/2$. 
The state of the qutrit evolves as $\rho\ts{d,1} \to \rho\ts{d,2}$, and $\rho\ts{d,2}$ can be calculated following the same procedure as $\rho \ts{d,1}$: 
\begin{align}
    \rho\ts{d,2}
    &=
        \begin{bmatrix}
            p_2 + \delph{2} + \delpc{2} & 0 & \mathcal{C}\ts{i} + \mathcal{C}\ts{ii} \\
            0 & p_1 + \delph{1} + \delpc{1} & 0 \\
            \mathcal{C}\ts{i}^* + \mathcal{C}\ts{ii}^* & 0 & 1-p_1-p_2-\delph{1} - \delph{2} -\delpc{1} - \delpc{2}
        \end{bmatrix}
        + 
        \mathcal{O}(\lambda^4)\,,
\end{align}
\end{widetext}
where $\delpc{2}, \delpc{1}, \mathcal{C}\ts{ii}$, and $\resc{\pm \Omega\ssc{ii}_{ij}}$ take the same form as Eqs.~\eqref{eq:delta p2H}, \eqref{eq:delta p1H}, \eqref{eq:hot coherence term}, and \eqref{eq:response hot}, respectively, replacing the subscript `I' with `II'. 
Therefore, the heat and work exchanged during this stroke are 
\begin{equation}
    \braket{Q_4}
    =
        (\delpc{2} +\delpc{1}) \Omega_{01}\ssc{ii}
        + 
        \delpc{2} \Omega_{12}\ssc{ii}\,,
        \quad 
    \braket{W_4}
    =
        0\,.
\end{equation}

Finally, to close the cycle, we impose the condition $\rho\ts{d,2}=\rho\ts{d,0}$, which implies
\begin{subequations}
\begin{align}
    &\delph{1} + \delpc{1} 
    =
        0\,, \label{close condition 0} \\
    &\delph{2} 
    + 
    \delpc{2} = 0\,, \label{close condition 2} \\
    &\mathcal{C}\ts{i} + \mathcal{C}\ts{ii}
        =0\,. \label{close condition offdiag}
\end{align} \label{eq:cycle conditions}
\end{subequations}
As we will discuss in more detail in Sec.~\ref{sec:conclusion}, it should be noted that, in order to close the cycle, these coherence terms $\mathcal{C}\ts{i}$ and $\mathcal{C}\ts{ii}$ must be eliminated, which may require additional energy. 
Although such energy must be accounted for to compute the extracted work properly, its estimation is complicated. 
Hence, we focus solely on Alicki’s definition of work \cite{R.Alicki_1979} [see, e.g., Eq.~\eqref{eq:W1}] and examine the PWC.

\section{Result}\label{sec:results}
\subsection{General expression for the PWC}

From (step-1) to (step-4), the total amount of extracted work $\wext$ reads 
\begin{align}
    \wext 
    &= 
        -(\langle W_1 \rangle + \langle W_3 \rangle) \notag \\
    &=
        (\delph{2} +\delph{1}) \Delta \Omega_{01} 
        + 
        \delph{2} \Delta \Omega_{12}\,,
    \label{eq:Wext}
\end{align}
and the positive work condition (PWC) is expressed as $\wext>0$. 
Furthermore, we impose \eqref{eq:cycle conditions} to close the thermodynamic cycle. 
By following the calculation given in Appendix~\ref{app:derivation of PWC}, the PWC for a qutrit-type RQOE reads 
\begin{align}
    \mathcal{A}(\Omega_{01})
    S(\Omega_{01}) \Delta \Omega_{01}
    +
    \mathcal{A}(\Omega_{21})
    S(\Omega_{21}) \Delta \Omega_{21} > 0\,, \label{eq:PWC qutrit general}
\end{align}
where $\Omega_{21}\ssc{i,ii}=-\Omega_{12}\ssc{i,ii}$, $\Delta \Omega_{21}=-\Delta \Omega_{12}$, and 
\begin{subequations}
\begin{align}
    \mathcal{A}(\Omega_{ij})
    &\equiv 
        \mathcal{A}(\Omega\ssc{i}_{ij},\Omega\ssc{ii}_{ij}) \notag \\
    &\coloneqq
        \kagikako{
            \dfrac{1}{ \sigma\ts{i} \resh{\Omega\ssc{i}_{ij}} }
            +
            \dfrac{1}{ \sigma\ts{ii} \resc{\Omega\ssc{ii}_{ij}} }
        }^{-1}
        (>0)\,, \\
    S(\Omega_{ij})
    &\equiv 
        S(\Omega\ssc{i}_{ij}, \Omega\ssc{ii}_{ij})
    \coloneqq
        e^{ \Omega\ssc{ii}_{ij}/\tempsec{\Omega\ssc{ii}_{ij}}  }
        - 
        e^{ \Omega\ssc{i}_{ij}/\tempfirst{\Omega\ssc{i}_{ij}}  }\,. \label{eq:S Omega}
\end{align}
\end{subequations}
Here, $T\sst{eff}_j(\Omega)$, $j\in \{ \text{I}, \text{II} \}$, is an \textit{effective temperature with respect to the two levels gapped by $\Omega$} defined as \cite{Gallock2023Otto}
\begin{align}
    \dfrac{1}{T\sst{eff}_j (\Omega)}
    \coloneqq
        \dfrac{1}{\Omega}
        \ln \dfrac{ \mathcal{F}_j(-\Omega) }{ \mathcal{F}_j(\Omega) }\,. \label{eq:def of eff T}
\end{align}
The effective temperature defined above is well-posed for two-level quantum systems in the sense that it reduces to the bona fide temperature of the environment, $T\sst{eff} \to T$, if the environment is in the thermal state and the two-level system reaches thermal equilibrium. 
To be precise, the effective temperature becomes the bona fide temperature if the pullback of the Wightman function satisfies the KMS conditions and the adiabatic long-interaction limit is taken \cite{Fewster.Waiting.Unruh, Garay2016anti-unruh}.
We note, however, that for general multi-level quantum systems with inhomogeneous energy gaps, the effective temperature is not unique as it depends on which pair of energy eigenstates we are looking at. 
For our qutrit system, we have two notions of effective temperature, $T\sst{eff}_j(\Omega_{01})$ and $T\sst{eff}_j(\Omega_{12})$, and they are inequivalent unless the energy gaps are equal \cite{Quan.QHE.2007}. 
For this reason, we need to specify a pair of energy eigenstates to talk about the effective temperature. 
Nevertheless, if the multilevel system reaches thermal equilibrium, every effective temperature coincides with the environment's bona fide temperature.

We note that the notion of ``closing a cycle'' is subtle in quantum thermodynamics. 
In classical and quantum thermodynamics, it is typically assumed that heat baths remain constant and thereby, one closes a cycle by bringing back the state of a working substance to its initial state. 
However, if one considers environments that are subject to change, then closing a cycle must mean bringing back the whole system to their initial state. 
This is crucial in quantum thermodynamics, especially when a working substance entangles with the environment, or when one is interested in the non-Markovian behavior of the environment. 
In our case, the qutrit is entangled with the quantum field, and so we need to add some extra work to disentangle them at the end. 
Although we impose the closing cycle condition only for the working substance in this work, the state of the environment and the correlation developed should be taken into account if one wishes to investigate work extraction in the strict sense.

\begin{figure}[t]
    \centering
\includegraphics[width=1.0\linewidth]{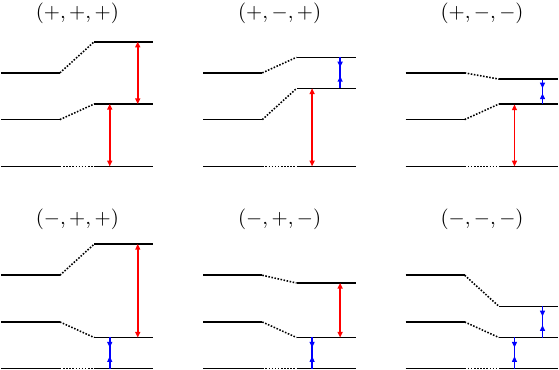}
    \caption{All six cases for $(\Delta \Omega_{01}, \Delta \Omega_{12}, \Delta \Omega_{02})$. }
    \label{fig:gap change all cases}
\end{figure}

\begin{figure*}[t]
    \centering
\includegraphics[width=0.9\linewidth]{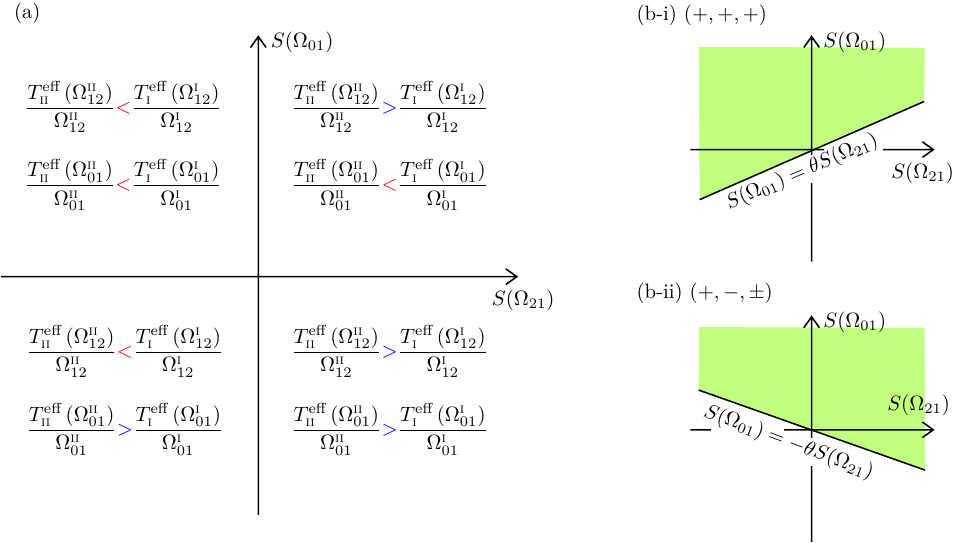}
    \caption{Visualization of the qutrit PWC in the $S(\Omega_{21})$-$S(\Omega_{01})$ plane. 
    (a) Each quadrant can be expressed in terms of the effective temperatures. 
    (b) The green areas represent the regions in the $S(\Omega_{21})$-$S(\Omega_{01})$ plane that satisfies the qutrit PWC. 
    }
    \label{fig:PWC patterns}
\end{figure*}

\subsection{Case study of various energy gaps}

Analyzing QOEs using two-level systems is very simple since there is only one energy gap. 
In particular, the qubit-PWC in terms of $S(\Omega)$ defined in \eqref{eq:S Omega} can be expressed as $S(\Omega)>0$ if the energy gap is assumed to be enlarged during the first adiabatic process. 
In contrast to the qubit-type QOE, qutrit-type QOEs have various ways to manipulate each energy gap. 
For example, during the first adiabatic process, one could enlarge the gap $\Omega_{01}$ between the states $\ket{e_0}$ and $\ket{e_1}$ while reducing the gap $\Omega_{12}$ between $\ket{e_1}$ and $\ket{e_2}$. 
Moreover, the qutrit-PWC \eqref{eq:PWC qutrit general} consists of $S(\Omega_{01})$ and $S(\Omega_{21})$ influencing each other. 
Our aim here is to examine each scenario and compare them to the well-known PWC for the qubit-type QOE.

To begin with, let us consider the simplest scenario of an evenly gapped qutrit, $\Omega_{01}=\Omega_{12} \equiv \Omega$. 
It is straightforward to check that the PWC \eqref{eq:PWC qutrit general} in this case is written as $S(\Omega) + S(-\Omega)>0$ and is equivalent to that for the two-level system: 
\begin{align}
    \dfrac{ \tempsec{\Omega\ssc{ii}} }{\Omega\ssc{ii}}
    <
    \dfrac{ \tempfirst{\Omega\ssc{i}} }{\Omega\ssc{i}}\,,
\end{align}
where we used the fact that $T\sst{eff}_j(-\Omega) = T\sst{eff}_j(\Omega)$ and $\cosh x > \cosh y \Rightarrow x > y (>0)$. 
As described in \cite{Gallock2023Otto}, if the pullback of the Wightman function satisfies the KMS condition and the detector interacts with the field for a sufficiently long time, then the effective temperatures become the bona fide temperatures of the field, yielding the well-known PWC in the literature \cite{Feldmann.QOE.2000, Kieu.secondlaw.demon.otto}, $T\ts{ii}/\Omega\ssc{ii} < T\ts{i}/\Omega\ssc{i}$.

Let us now examine various cases where $\Omega_{01}\neq \Omega_{12}$. 
We can categorize the three-level quantum Otto cycle by the sign of $\Delta \Omega_{01}, \Delta \Omega_{12}$, and $\Delta \Omega_{02}$. 
Recalling that $\Omega_{02}=\Omega_{01} + \Omega_{12}$ and so $\Delta \Omega_{02}= \Delta \Omega_{01} + \Delta \Omega_{12}$, we have six scenarios to consider. 
For an explanation, consider a triple $(\Delta \Omega_{01}, \Delta \Omega_{12}, \Delta \Omega_{02})$, where each slot can be either $+$ or $-$ depending on the sign. 
For example, $(+,-,+)$ means $\Delta \Omega_{01} > 0$, $\Delta \Omega_{12} < 0$, and $\Delta \Omega_{02} > 0$. 
Noticing that it is impossible to have $(+,+,-)$ and $(-,-,+)$ due to the constraint $\Delta \Omega_{02}= \Delta \Omega_{01} + \Delta \Omega_{12}$, we have the following six cases: $(+,+,+)$, $(+,-,+)$, $(-,+,+)$, $(+,-,-)$, $(-,+,-)$, and $(-,-,-)$. 
These are depicted in Fig.~\ref{fig:gap change all cases}. 
Furthermore, swapping $\Delta \Omega_{ij} \leftrightarrow -\Delta \Omega_{ij}$ and the order of the isochoric strokes $\text{I} \leftrightarrow \text{II}$ yields an equivalent cycle with a different starting point. 
For this reason, we mainly focus on the three cases, $(+,+,+)$ and $(+, -, \pm)$.

\subsubsection{Case: $(+,+,+)$}
This is the case where both the energy gaps $\Omega_{01}$ and $\Omega_{12}$ are enlarged during the first adiabatic process: $\Delta \Omega\ts{01}>0$, $\Delta \Omega\ts{12}>0$. 
Then, the qutrit-PWC \eqref{eq:PWC qutrit general} can be written as 
\begin{align}
    &S(\Omega_{01})
    >
    \theta S(\Omega_{21})
    \,,  \label{eq:+++}
\end{align}
where 
\begin{align}
    \theta
    &\equiv 
        \theta(\Omega_{01}\ssc{i}, \Omega_{01}\ssc{ii}, \Omega_{12}\ssc{i}, \Omega_{12}\ssc{ii})
    \coloneqq
        \dfrac{ \mathcal{A}(\Omega_{21}) }{ \mathcal{A}(\Omega_{01}) }
        \left| 
            \dfrac{\Delta \Omega_{12}}{ \Delta \Omega_{01} }
        \right| >0
        \,, \label{eq:theta}
\end{align}
and we have used $\Omega_{21}=-\Omega_{12}$. 
Notice that the expression \eqref{eq:+++} for the PWC is more complicated than that for qubits: $S(\Omega)>0$. 
While the simplicity of the qubit PWC enables us to express it in terms of temperatures, $\tempsec{\Omega\ssc{ii}} /\Omega\ssc{ii} < \tempfirst{\Omega\ssc{i}}/ \Omega\ssc{i}$, this is no longer the case for qutrits.

For a further analysis, consider an $S(\Omega_{21})$-$S(\Omega_{01})$ plane depicted in Fig.~\ref{fig:PWC patterns}. 
Each quadrant can be characterized in terms of the qubit-type PWC for each pair of energy levels. 
For example, in Fig.~\ref{fig:PWC patterns}(a), the first quadrant corresponds to the region with $S(\Omega_{21})>0$ and $S(\Omega_{01})>0$, which reduce to 
\begin{align}
    \dfrac{ \tempsec{\Omega_{12}\ssc{ii} } }{ \Omega_{12}\ssc{ii} }
    >
        \dfrac{ \tempfirst{\Omega_{12}\ssc{i}} }{ \Omega_{12}\ssc{i} }\,,
    \quad
    \dfrac{ \tempsec{\Omega_{01}\ssc{ii} } }{ \Omega_{01}\ssc{ii} }
    <
        \dfrac{ \tempfirst{\Omega_{01}\ssc{i}} }{ \Omega_{01}\ssc{i} }\,. \notag 
\end{align}
The qutrit-PWC \eqref{eq:+++} for $(+,+,+)$ can be represented as a green region in Fig.~\ref{fig:PWC patterns}(b-i). 
The boundary of the PWC in this diagram is a straight line $S(\Omega_{01})= \theta S( \Omega_{21} )$ passing through the origin of the plane. 
The slope of such a line depends on the parameters that determine $\theta=\theta(\Omega_{01}\ssc{i}, \Omega_{01}\ssc{ii}, \Omega_{12}\ssc{i}, \Omega_{12}\ssc{ii})$.

The diagram allows us to compare the qutrit-QOEs to those for qubits. 
Recall that the PWC for qubits is expressed as $S(\Omega)>0$, or equivalently $\tempsec{\Omega\ssc{ii}} /\Omega\ssc{ii} < \tempfirst{\Omega\ssc{i}}/ \Omega\ssc{i}$, when $\Delta \Omega>0$. 
For qutrits, the qubit-PWC becomes a sufficient condition as it can be seen from the second quadrant in Fig.~\ref{fig:PWC patterns}(b-i). 
Namely, if both pairs of energy eigenstates with energy gap $\Omega_{01}$ and $\Omega_{12}$ satisfy the qubit-PWC, then a qutrit can extract work. 
This is consistent with the previous research \cite{QHE.Quan.qutrit}, where the qutrit-PWC is shown to be a looser condition than that of qubits when the temperature of a hot reservoir is very high. 
However, our result adds more to the prior research; 
We find that qutrits can also extract work even if one of the pairs of energy gaps violates the qubit-PWC. 
This can be observed from the first and the third quadrants in Fig.~\ref{fig:PWC patterns}(b-i). 
For example, the green region in the first quadrant tells us that qutrits can extract positive work even when the pair $\ket{e_1}$ and $\ket{e_2}$ does not satisfy the qubit-PWC. 
It should be noted that the set of parameters that realizes such a scenario is restricted. 
Moreover, one can choose the set of parameters such that the boundary of the qutrit-PWC, $S(\Omega_{01})= \theta S( \Omega_{21} )$, is vertical or horizontal. 
These correspond to the scenarios where one of the energy gaps is fixed throughout the cycle, and the qutrit-PWC reduces to the qubit-PWC for the energy gap that is not fixed.

\subsubsection{Case: $(+, - , \pm)$}

Next, we focus on the case where the energy gap between $\ket{e_0}$ and $\ket{e_1}$ increases ($\Delta \Omega_{01}>0$) while that of the upper levels decreases ($\Delta \Omega_{12}<0$) during the first adiabatic process. 
Depending on the amount of $\Delta \Omega_{01}$ and $\Delta \Omega_{12}$, the change in the overall energy gap $\Delta \Omega_{02}$ can either increase or decrease.

For both $(+,-,+)$ and $(+,-,-)$, the qutrit-PWC in \eqref{eq:PWC qutrit general} can be expressed as 
\begin{align}
    S(\Omega_{01}) > -\theta S(\Omega_{21})\,, \label{eq:PWC +-+}
\end{align}
where $\theta>0$ is defined in \eqref{eq:theta}. 
We again note that $\theta$ and therefore the slope of the boundary described by $S(\Omega_{01}) = -\theta S(\Omega_{21})$ depend on a set of energy gaps $\Omega_{01}\ssc{i}$, $\Omega_{01}\ssc{ii}$, $\Omega_{12}\ssc{i}$, and $\Omega_{12}\ssc{ii}$.

The green area in Fig.~\ref{fig:PWC patterns}(b-ii) shows the region in the $S(\Omega_{21})$-$S(\Omega_{01})$ plane satisfying the PWC \eqref{eq:PWC +-+} for $(+,-,\pm)$. 
While $(+,-,+)$ and $(+,-,-)$ generally have different $\theta$, they both share generic features. 
Unlike the $(+,+,+)$ case, the $(+,-,\pm)$ scenario cannot extract work in the third quadrant, where the qubit-PWC holds for $\Omega_{12}$ but is violated for $\Omega_{01}$. 
Moreover, even when the qubit-PWC is satisfied for both $\Omega_{01}$ and $\Omega_{12}$, work extraction is prohibited for a certain choice of parameters, as indicated in the second quadrant in Fig.~\ref{fig:PWC patterns}(b-ii). 
Instead, it is possible to extract work from the quantum field even when the qubit-PWC violated for both $\Omega_{01}$ and $\Omega_{12}$ [the fourth quadrant in Fig.~\ref{fig:PWC patterns}(b-ii)].

\section{Conclusion and Discussion}\label{sec:conclusion}

In this paper, we considered an RQOE that employs a qutrit-type UDW detector as its working substance and derived the PWC \eqref{eq:PWC qutrit general} in terms of the response functions and the effective temperature. 
Here, the effective temperature is understood as the temperature perceived by a specific pair of energy eigenstates of the qutrit. 
Thus, unless the energy levels are equally gapped, each level experiences a different temperature. 
Nevertheless, once the qutrit reaches thermal equilibrium, these effective temperatures merge into a unique bona fide temperature of the field. 
Precisely, this bona fide temperature emerges when the pullback of the Wightman function of the quantum field satisfies the KMS conditions and the qutrit interacts adiabatically with the field for an infinitely long time \cite{Fewster.Waiting.Unruh, Garay2016anti-unruh}.

While the well-known qubit-PWC can be expressed simply in terms of the ratio of the effective temperature and the energy gap \cite{Feldmann.QOE.2000, Kieu.heat.engine.2006, Gallock2023Otto}, the qutrit-PWC exhibits influences among the three levels, making the expression complicated. 
Moreover, qutrits allow for various manipulations of the energy gaps. 
This leads to a rich variety of qutrit-PWC.

As summarized in Fig.~\ref{fig:PWC patterns}, the qutrit-PWC can be compared to the qubit-PWC for each pair of energy eigenstates. 
As reported in \cite{QHE.Quan.qutrit}, we observed that the qutrit-PWC is looser than the qubit-PWC when all the energy gaps increase during the first adiabatic process. 
Nevertheless, we additionally found that qutrits can still extract work even when one pair of energy eigenstates fails to satisfy the qubit-PWC, though the set of parameters that realizes such work extraction is restricted. 
These aspects are less studied in the literature.

While qubits are the simplest working substance to deal with, they often exhibit exceptional characteristics that are absent in general multilevel systems. 
Examining qutrits provides us insight into the quantum thermodynamic properties of generic quantum systems. 
In this regard, we close this section by addressing several subtleties that multilevel systems encounter in QHEs.

The first subtlety is the existence of an off-diagonal coherence term in the density matrix. 
For a given QOE, the density matrix of the working substance does not have off-diagonal elements provided that the system thermalizes at each stroke. 
Even when thermalization does not occur (e.g., due to a finite-time interaction), a qubit does not acquire an off-diagonal coherence term at the leading order in the coupling constant within perturbation theory. 
However, three-level systems inevitably acquire off-diagonal elements when thermalization is not guaranteed. 
This is problematic for closing a cycle, as the newly acquired coherence term must be reset at the end, which requires additional energy.

Another subtlety, related to the aforementioned coherence terms, is the definition of work and heat in quantum thermodynamics.  
In the literature, Alicki's definition of work \cite{R.Alicki_1979}, which is essentially the expectation value of the change in the system Hamiltonian $\braket{W}=\Tr[\rho \Delta \hat H]$, is commonly employed.  
Under this definition, the off-diagonal coherence terms in the system's density matrix $\rho$ are not accounted for in the work.  
Nevertheless, several approaches exist for handling coherence terms.  
For example, in \cite{shi2020quantum.coherence}, the contributions of coherence are explicitly incorporated into the definitions of work and heat, and these contributions cancel out at the end of the cycle, yielding a first law of thermodynamics that excludes coherence terms.  
In other words, the coherence received from the heat bath is directly returned as work.  
When applied to our model, this means that coherence naturally vanishes after one complete cycle.  
Furthermore, the presence of coherence allows for additional work extraction, thereby relaxing the PWC compared to the case without coherence.  
However, there remains ongoing debate regarding whether coherence should be included in the definitions of work and heat.  
For example, Ref.~\cite{B.Lima} defines the first law of thermodynamics to include a coherence term, instead of modifying the definitions of work and heat as in \cite{shi2020quantum.coherence}. 
In such a case, the coherence term in the first law must be eliminated at the end of the cycle if the coherence is initially absent in the system. 
In any case, future research in both RQI and quantum thermodynamics must place great emphasis on the existence of off-diagonal coherence terms. 



\acknowledgments{
This work was partially supported by Grant-in-Aid for Research Activity Start-up (Grant No. JP24K22862) and by Grant-in-Aid for JSPS Fellows Grant Number JP25KJ0048. 
}

\begin{widetext}
\appendix
\section{Derivation of the PWC}\label{app:derivation of PWC}

In this section, we derive our qutrit-PWC given in \eqref{eq:PWC qutrit general}. 
Reminding ourselves that the work extracted \eqref{eq:Wext} is 
\begin{align}
    \wext 
    &=
        (\delph{2} +\delph{1}) \Delta \Omega_{01} 
        + 
        \delph{2} \Delta \Omega_{12}\,, \notag 
\end{align}
we impose the cyclicity conditions \eqref{eq:cycle conditions} and determine the initial population of each energy level, $p_1$ and $p_2$, in terms of the response functions. 
Substituting $\delph{2}$ given in \eqref{eq:delta p2H} and $\delpc{2}$ [\eqref{eq:delta p2H} with $\text{I}\to \text{II}$] into the condition \eqref{close condition 2}, we obtain a formula connecting $p_2$ and $p_1$: 
\begin{subequations}
\begin{align}
    p_1
    &=
        \Xi^{-1}
        [ \sigma\ts{i} \resh{\Omega_{01}\ssc{i}} + \sigma\ts{ii} \resc{\Omega_{01}\ssc{ii}} ] [ \sigma\ts{i} \resh{-\Omega_{12}\ssc{i}} + \sigma\ts{ii} \resc{-\Omega_{12}\ssc{ii}} ]\,, \\
    p_2
    &=
        \Xi^{-1}
        [ \sigma\ts{i} \resh{\Omega_{01}\ssc{i}} + \sigma\ts{ii} \resc{\Omega_{01}\ssc{ii}} ] [ \sigma\ts{i} \resh{\Omega_{12}\ssc{i}} + \sigma\ts{ii} \resc{\Omega_{12}\ssc{ii}} ]\,, \\
    \Xi
    &\coloneqq
        \sigma\ts{ii}^2 \resc{\Omega_{01}\ssc{ii}} \resc{\Omega_{12}\ssc{ii}}
        +
        \sigma\ts{ii}^2 \resc{\Omega_{01}\ssc{ii}} 
        \resc{-\Omega_{12}\ssc{ii}}
        +
        \sigma\ts{ii}^2 \resc{-\Omega_{01}\ssc{ii}} 
        \resc{-\Omega_{12}\ssc{ii}}
        +
        \sigma\ts{i} \sigma\ts{ii} 
        \resh{\Omega_{01}\ssc{i}} 
        \resc{\Omega_{12}\ssc{ii}} \notag \\
        &\quad
        +
        \sigma\ts{i} \sigma\ts{ii} 
        \resh{\Omega_{01}\ssc{i}} 
        \resc{-\Omega_{12}\ssc{ii}}
        +
        \sigma\ts{i} \sigma\ts{ii} 
        \resh{\Omega_{12}\ssc{i}} 
        \resc{\Omega_{01}\ssc{ii}}
        +
        \sigma\ts{i} \sigma\ts{ii} 
        \resh{-\Omega_{01}\ssc{i}} 
        \resc{-\Omega_{12}\ssc{ii}}
        +
        \sigma\ts{i} \sigma\ts{ii} 
        \resh{-\Omega_{12}\ssc{i}} 
        \resc{\Omega_{01}\ssc{ii}} \notag \\
        &\quad
        +
        \sigma\ts{i} \sigma\ts{ii} 
        \resh{-\Omega_{12}\ssc{i}} 
        \resc{-\Omega_{01}\ssc{ii}}
        +
        \sigma\ts{i}^2
        \resh{\Omega_{01}\ssc{i}} 
        \resh{\Omega_{12}\ssc{i}}
        +
        \sigma\ts{i}^2
        \resh{\Omega_{01}\ssc{i}} 
        \resh{-\Omega_{12}\ssc{i}}
        +
        +
        \sigma\ts{i}^2
        \resh{-\Omega_{01}\ssc{i}} 
        \resh{-\Omega_{12}\ssc{i}}\,.
\end{align}
\end{subequations}
Substituting these $p_1$ and $p_2$ into $\delph{1}$ and $\delph{2}$ in \eqref{eq:delta p2H} and \eqref{eq:delta p1H} allows us to express $\delph{1}+\delph{2}$ and $\delph{2}$ in terms of the response functions: 
\begin{align}
    \delph{1}+\delph{2}
    &=
        \lambda^2 \sigma\ts{i} \sigma\ts{ii} \Gamma^{-1}
        [ \sigma\ts{i} \resh{-\Omega_{12}\ssc{i}} + \sigma\ts{ii} \resc{-\Omega_{12}\ssc{ii}} ]
        [ \resh{\Omega_{01}\ssc{i}} \resc{-\Omega_{01}\ssc{ii}} - \resh{-\Omega_{01}\ssc{i}} \resc{\Omega_{01}\ssc{ii}} ]\,, \\
    \delph{2}
    &=
        \lambda^2 \sigma\ts{i} \sigma\ts{ii} \Gamma^{-1}
        [ \sigma\ts{i} \resh{\Omega_{01}\ssc{i}} + \sigma\ts{ii} \resc{\Omega_{01}\ssc{ii}} ]
        [ \resh{\Omega_{12}\ssc{i}} \resc{-\Omega_{12}\ssc{ii}} - \resh{-\Omega_{12}\ssc{i}} \resc{\Omega_{12}\ssc{ii}} ]\,,
\end{align}
where $\Gamma>0$ is a positive number composed of response functions. 
We then substitute these into the qutrit-PWC $\wext>0$. 
By introducing the effective temperature defined in \eqref{eq:def of eff T}, we obtain our main result in \eqref{eq:PWC qutrit general}. 

\end{widetext}

\bibliography{ref}

\begin{thebibliography}{54}%
\makeatletter
\providecommand \@ifxundefined [1]{%
 \@ifx{#1\undefined}
}%
\providecommand \@ifnum [1]{%
 \ifnum #1\expandafter \@firstoftwo
 \else \expandafter \@secondoftwo
 \fi
}%
\providecommand \@ifx [1]{%
 \ifx #1\expandafter \@firstoftwo
 \else \expandafter \@secondoftwo
 \fi
}%
\providecommand \natexlab [1]{#1}%
\providecommand \enquote  [1]{``#1''}%
\providecommand \bibnamefont  [1]{#1}%
\providecommand \bibfnamefont [1]{#1}%
\providecommand \citenamefont [1]{#1}%
\providecommand \href@noop [0]{\@secondoftwo}%
\providecommand \href [0]{\begingroup \@sanitize@url \@href}%
\providecommand \@href[1]{\@@startlink{#1}\@@href}%
\providecommand \@@href[1]{\endgroup#1\@@endlink}%
\providecommand \@sanitize@url [0]{\catcode `\\12\catcode `\$12\catcode `\&12\catcode `\#12\catcode `\^12\catcode `\_12\catcode `\%12\relax}%
\providecommand \@@startlink[1]{}%
\providecommand \@@endlink[0]{}%
\providecommand \url  [0]{\begingroup\@sanitize@url \@url }%
\providecommand \@url [1]{\endgroup\@href {#1}{\urlprefix }}%
\providecommand \urlprefix  [0]{URL }%
\providecommand \Eprint [0]{\href }%
\providecommand \doibase [0]{https://doi.org/}%
\providecommand \selectlanguage [0]{\@gobble}%
\providecommand \bibinfo  [0]{\@secondoftwo}%
\providecommand \bibfield  [0]{\@secondoftwo}%
\providecommand \translation [1]{[#1]}%
\providecommand \BibitemOpen [0]{}%
\providecommand \bibitemStop [0]{}%
\providecommand \bibitemNoStop [0]{.\EOS\space}%
\providecommand \EOS [0]{\spacefactor3000\relax}%
\providecommand \BibitemShut  [1]{\csname bibitem#1\endcsname}%
\let\auto@bib@innerbib\@empty
\bibitem [{\citenamefont {Vinjanampathy}\ and\ \citenamefont {Anders}(2016)}]{Vinjanampathy_2016}%
  \BibitemOpen
  \bibfield  {author} {\bibinfo {author} {\bibfnamefont {S.}~\bibnamefont {Vinjanampathy}}\ and\ \bibinfo {author} {\bibfnamefont {J.}~\bibnamefont {Anders}},\ }\bibfield  {title} {\bibinfo {title} {Quantum thermodynamics},\ }\href {https://doi.org/10.1080/00107514.2016.1201896} {\bibfield  {journal} {\bibinfo  {journal} {Contemporary Physics}\ }\textbf {\bibinfo {volume} {57}},\ \bibinfo {pages} {545–579} (\bibinfo {year} {2016})}\BibitemShut {NoStop}%
\bibitem [{\citenamefont {Goold}\ \emph {et~al.}(2016)\citenamefont {Goold}, \citenamefont {Huber}, \citenamefont {Riera}, \citenamefont {del Rio},\ and\ \citenamefont {Skrzypczyk}}]{Goold_2016}%
  \BibitemOpen
  \bibfield  {author} {\bibinfo {author} {\bibfnamefont {J.}~\bibnamefont {Goold}}, \bibinfo {author} {\bibfnamefont {M.}~\bibnamefont {Huber}}, \bibinfo {author} {\bibfnamefont {A.}~\bibnamefont {Riera}}, \bibinfo {author} {\bibfnamefont {L.}~\bibnamefont {del Rio}},\ and\ \bibinfo {author} {\bibfnamefont {P.}~\bibnamefont {Skrzypczyk}},\ }\bibfield  {title} {\bibinfo {title} {The role of quantum information in thermodynamics—a topical review},\ }\href {https://doi.org/10.1088/1751-8113/49/14/143001} {\bibfield  {journal} {\bibinfo  {journal} {Journal of Physics A: Mathematical and Theoretical}\ }\textbf {\bibinfo {volume} {49}},\ \bibinfo {pages} {143001} (\bibinfo {year} {2016})}\BibitemShut {NoStop}%
\bibitem [{\citenamefont {Gemmer}\ \emph {et~al.}(2009)\citenamefont {Gemmer}, \citenamefont {Michel},\ and\ \citenamefont {Mahler}}]{gemmer2009quantum}%
  \BibitemOpen
  \bibfield  {author} {\bibinfo {author} {\bibfnamefont {J.}~\bibnamefont {Gemmer}}, \bibinfo {author} {\bibfnamefont {M.}~\bibnamefont {Michel}},\ and\ \bibinfo {author} {\bibfnamefont {G.}~\bibnamefont {Mahler}},\ }\href {https://doi.org/10.1007/978-3-540-70510-9} {\emph {\bibinfo {title} {{Quantum thermodynamics: Emergence of thermodynamic behavior within composite quantum systems}}}},\ Vol.\ \bibinfo {volume} {784}\ (\bibinfo  {publisher} {Springer},\ \bibinfo {year} {2009})\BibitemShut {NoStop}%
\bibitem [{\citenamefont {Scovil}\ and\ \citenamefont {Schulz-DuBois}(1959)}]{Scovil.heat.engine.maser}%
  \BibitemOpen
  \bibfield  {author} {\bibinfo {author} {\bibfnamefont {H.~E.~D.}\ \bibnamefont {Scovil}}\ and\ \bibinfo {author} {\bibfnamefont {E.~O.}\ \bibnamefont {Schulz-DuBois}},\ }\bibfield  {title} {\bibinfo {title} {{Three-Level Masers as Heat Engines}},\ }\href {https://doi.org/10.1103/PhysRevLett.2.262} {\bibfield  {journal} {\bibinfo  {journal} {Phys. Rev. Lett.}\ }\textbf {\bibinfo {volume} {2}},\ \bibinfo {pages} {262} (\bibinfo {year} {1959})}\BibitemShut {NoStop}%
\bibitem [{\citenamefont {Feldmann}\ and\ \citenamefont {Kosloff}(2000)}]{Feldmann.QOE.2000}%
  \BibitemOpen
  \bibfield  {author} {\bibinfo {author} {\bibfnamefont {T.}~\bibnamefont {Feldmann}}\ and\ \bibinfo {author} {\bibfnamefont {R.}~\bibnamefont {Kosloff}},\ }\bibfield  {title} {\bibinfo {title} {Performance of discrete heat engines and heat pumps in finite time},\ }\href {https://doi.org/10.1103/PhysRevE.61.4774} {\bibfield  {journal} {\bibinfo  {journal} {Phys. Rev. E}\ }\textbf {\bibinfo {volume} {61}},\ \bibinfo {pages} {4774} (\bibinfo {year} {2000})}\BibitemShut {NoStop}%
\bibitem [{\citenamefont {Kieu}(2004)}]{Kieu.secondlaw.demon.otto}%
  \BibitemOpen
  \bibfield  {author} {\bibinfo {author} {\bibfnamefont {T.~D.}\ \bibnamefont {Kieu}},\ }\bibfield  {title} {\bibinfo {title} {{The Second Law, Maxwell's Demon, and Work Derivable from Quantum Heat Engines}},\ }\href {https://doi.org/10.1103/PhysRevLett.93.140403} {\bibfield  {journal} {\bibinfo  {journal} {Phys. Rev. Lett.}\ }\textbf {\bibinfo {volume} {93}},\ \bibinfo {pages} {140403} (\bibinfo {year} {2004})}\BibitemShut {NoStop}%
\bibitem [{\citenamefont {Kieu}(2006)}]{Kieu.heat.engine.2006}%
  \BibitemOpen
  \bibfield  {author} {\bibinfo {author} {\bibfnamefont {T.~D.}\ \bibnamefont {Kieu}},\ }\bibfield  {title} {\bibinfo {title} {{Quantum heat engines, the second law and Maxwell's daemon}},\ }\href {https://doi.org/https://doi.org/10.1140/epjd/e2006-00075-5} {\bibfield  {journal} {\bibinfo  {journal} {Eur. Phys. J. D}\ }\textbf {\bibinfo {volume} {39}},\ \bibinfo {pages} {115} (\bibinfo {year} {2006})}\BibitemShut {NoStop}%
\bibitem [{\citenamefont {Rostovtsev}\ \emph {et~al.}(2003)\citenamefont {Rostovtsev}, \citenamefont {Matsko}, \citenamefont {Nayak}, \citenamefont {Zubairy},\ and\ \citenamefont {Scully}}]{Rostovtsev.Otto.2003}%
  \BibitemOpen
  \bibfield  {author} {\bibinfo {author} {\bibfnamefont {Y.~V.}\ \bibnamefont {Rostovtsev}}, \bibinfo {author} {\bibfnamefont {A.~B.}\ \bibnamefont {Matsko}}, \bibinfo {author} {\bibfnamefont {N.}~\bibnamefont {Nayak}}, \bibinfo {author} {\bibfnamefont {M.~S.}\ \bibnamefont {Zubairy}},\ and\ \bibinfo {author} {\bibfnamefont {M.~O.}\ \bibnamefont {Scully}},\ }\bibfield  {title} {\bibinfo {title} {{Improving engine efficiency by extracting laser energy from hot exhaust gas}},\ }\href {https://doi.org/10.1103/PhysRevA.67.053811} {\bibfield  {journal} {\bibinfo  {journal} {Phys. Rev. A}\ }\textbf {\bibinfo {volume} {67}},\ \bibinfo {pages} {053811} (\bibinfo {year} {2003})}\BibitemShut {NoStop}%
\bibitem [{\citenamefont {Quan}\ \emph {et~al.}(2005)\citenamefont {Quan}, \citenamefont {Zhang},\ and\ \citenamefont {Sun}}]{QHE.Quan.qutrit}%
  \BibitemOpen
  \bibfield  {author} {\bibinfo {author} {\bibfnamefont {H.~T.}\ \bibnamefont {Quan}}, \bibinfo {author} {\bibfnamefont {P.}~\bibnamefont {Zhang}},\ and\ \bibinfo {author} {\bibfnamefont {C.~P.}\ \bibnamefont {Sun}},\ }\bibfield  {title} {\bibinfo {title} {Quantum heat engine with multilevel quantum systems},\ }\href {https://doi.org/10.1103/PhysRevE.72.056110} {\bibfield  {journal} {\bibinfo  {journal} {Phys. Rev. E}\ }\textbf {\bibinfo {volume} {72}},\ \bibinfo {pages} {056110} (\bibinfo {year} {2005})}\BibitemShut {NoStop}%
\bibitem [{\citenamefont {Quan}\ \emph {et~al.}(2007)\citenamefont {Quan}, \citenamefont {Liu}, \citenamefont {Sun},\ and\ \citenamefont {Nori}}]{Quan.QHE.2007}%
  \BibitemOpen
  \bibfield  {author} {\bibinfo {author} {\bibfnamefont {H.~T.}\ \bibnamefont {Quan}}, \bibinfo {author} {\bibfnamefont {Y.-x.}\ \bibnamefont {Liu}}, \bibinfo {author} {\bibfnamefont {C.~P.}\ \bibnamefont {Sun}},\ and\ \bibinfo {author} {\bibfnamefont {F.}~\bibnamefont {Nori}},\ }\bibfield  {title} {\bibinfo {title} {{Quantum thermodynamic cycles and quantum heat engines}},\ }\href {https://doi.org/10.1103/PhysRevE.76.031105} {\bibfield  {journal} {\bibinfo  {journal} {Phys. Rev. E}\ }\textbf {\bibinfo {volume} {76}},\ \bibinfo {pages} {031105} (\bibinfo {year} {2007})}\BibitemShut {NoStop}%
\bibitem [{\citenamefont {Scully}\ \emph {et~al.}(2003)\citenamefont {Scully}, \citenamefont {Zubairy}, \citenamefont {Agarwal},\ and\ \citenamefont {Walther}}]{Marlan_2003}%
  \BibitemOpen
  \bibfield  {author} {\bibinfo {author} {\bibfnamefont {M.~O.}\ \bibnamefont {Scully}}, \bibinfo {author} {\bibfnamefont {M.~S.}\ \bibnamefont {Zubairy}}, \bibinfo {author} {\bibfnamefont {G.~S.}\ \bibnamefont {Agarwal}},\ and\ \bibinfo {author} {\bibfnamefont {H.}~\bibnamefont {Walther}},\ }\bibfield  {title} {\bibinfo {title} {Extracting work from a single heat bath via vanishing quantum coherence},\ }\href {https://doi.org/10.1126/science.1078955} {\bibfield  {journal} {\bibinfo  {journal} {Science}\ }\textbf {\bibinfo {volume} {299}},\ \bibinfo {pages} {862} (\bibinfo {year} {2003})},\ \Eprint {https://arxiv.org/abs/https://www.science.org/doi/pdf/10.1126/science.1078955} {https://www.science.org/doi/pdf/10.1126/science.1078955} \BibitemShut {NoStop}%
\bibitem [{\citenamefont {Oppenheim}\ \emph {et~al.}(2002)\citenamefont {Oppenheim}, \citenamefont {Horodecki}, \citenamefont {Horodecki},\ and\ \citenamefont {Horodecki}}]{PhysRevLett.89.180402}%
  \BibitemOpen
  \bibfield  {author} {\bibinfo {author} {\bibfnamefont {J.}~\bibnamefont {Oppenheim}}, \bibinfo {author} {\bibfnamefont {M.}~\bibnamefont {Horodecki}}, \bibinfo {author} {\bibfnamefont {P.}~\bibnamefont {Horodecki}},\ and\ \bibinfo {author} {\bibfnamefont {R.}~\bibnamefont {Horodecki}},\ }\bibfield  {title} {\bibinfo {title} {Thermodynamical approach to quantifying quantum correlations},\ }\href {https://doi.org/10.1103/PhysRevLett.89.180402} {\bibfield  {journal} {\bibinfo  {journal} {Phys. Rev. Lett.}\ }\textbf {\bibinfo {volume} {89}},\ \bibinfo {pages} {180402} (\bibinfo {year} {2002})}\BibitemShut {NoStop}%
\bibitem [{\citenamefont {Dillenschneider}\ and\ \citenamefont {Lutz}(2009)}]{Dillenschneider_2009}%
  \BibitemOpen
  \bibfield  {author} {\bibinfo {author} {\bibfnamefont {R.}~\bibnamefont {Dillenschneider}}\ and\ \bibinfo {author} {\bibfnamefont {E.}~\bibnamefont {Lutz}},\ }\bibfield  {title} {\bibinfo {title} {Energetics of quantum correlations},\ }\href {https://doi.org/10.1209/0295-5075/88/50003} {\bibfield  {journal} {\bibinfo  {journal} {Europhysics Letters}\ }\textbf {\bibinfo {volume} {88}},\ \bibinfo {pages} {50003} (\bibinfo {year} {2009})}\BibitemShut {NoStop}%
\bibitem [{\citenamefont {Zhang}\ \emph {et~al.}(2007)\citenamefont {Zhang}, \citenamefont {Liu}, \citenamefont {Chen},\ and\ \citenamefont {Li}}]{Zhang_fourlevel_entangle_2007}%
  \BibitemOpen
  \bibfield  {author} {\bibinfo {author} {\bibfnamefont {T.}~\bibnamefont {Zhang}}, \bibinfo {author} {\bibfnamefont {W.-T.}\ \bibnamefont {Liu}}, \bibinfo {author} {\bibfnamefont {P.-X.}\ \bibnamefont {Chen}},\ and\ \bibinfo {author} {\bibfnamefont {C.-Z.}\ \bibnamefont {Li}},\ }\bibfield  {title} {\bibinfo {title} {{Four-level entangled quantum heat engines}},\ }\href {https://doi.org/10.1103/PhysRevA.75.062102} {\bibfield  {journal} {\bibinfo  {journal} {Phys. Rev. A}\ }\textbf {\bibinfo {volume} {75}},\ \bibinfo {pages} {062102} (\bibinfo {year} {2007})}\BibitemShut {NoStop}%
\bibitem [{\citenamefont {Uzdin}\ \emph {et~al.}(2015)\citenamefont {Uzdin}, \citenamefont {Levy},\ and\ \citenamefont {Kosloff}}]{Uzdin_2015}%
  \BibitemOpen
  \bibfield  {author} {\bibinfo {author} {\bibfnamefont {R.}~\bibnamefont {Uzdin}}, \bibinfo {author} {\bibfnamefont {A.}~\bibnamefont {Levy}},\ and\ \bibinfo {author} {\bibfnamefont {R.}~\bibnamefont {Kosloff}},\ }\bibfield  {title} {\bibinfo {title} {Equivalence of quantum heat machines, and quantum-thermodynamic signatures},\ }\href {https://doi.org/10.1103/PhysRevX.5.031044} {\bibfield  {journal} {\bibinfo  {journal} {Phys. Rev. X}\ }\textbf {\bibinfo {volume} {5}},\ \bibinfo {pages} {031044} (\bibinfo {year} {2015})}\BibitemShut {NoStop}%
\bibitem [{\citenamefont {{\"O}zdemir}\ and\ \citenamefont {M{\"u}stecaplio{\u{g}}lu}(2020)}]{ozdemir2020quantum}%
  \BibitemOpen
  \bibfield  {author} {\bibinfo {author} {\bibfnamefont {A.~T.}\ \bibnamefont {{\"O}zdemir}}\ and\ \bibinfo {author} {\bibfnamefont {{\"O}.~E.}\ \bibnamefont {M{\"u}stecaplio{\u{g}}lu}},\ }\bibfield  {title} {\bibinfo {title} {Quantum thermodynamics and quantum coherence engines},\ }\href {https://doi.org/10.3906/fiz-2009-12} {\bibfield  {journal} {\bibinfo  {journal} {Turkish Journal of Physics}\ }\textbf {\bibinfo {volume} {44}},\ \bibinfo {pages} {404} (\bibinfo {year} {2020})}\BibitemShut {NoStop}%
\bibitem [{\citenamefont {Yin}\ \emph {et~al.}(2020)\citenamefont {Yin}, \citenamefont {Chen}, \citenamefont {Wu},\ and\ \citenamefont {Ge}}]{yin2020work}%
  \BibitemOpen
  \bibfield  {author} {\bibinfo {author} {\bibfnamefont {Y.}~\bibnamefont {Yin}}, \bibinfo {author} {\bibfnamefont {L.}~\bibnamefont {Chen}}, \bibinfo {author} {\bibfnamefont {F.}~\bibnamefont {Wu}},\ and\ \bibinfo {author} {\bibfnamefont {Y.}~\bibnamefont {Ge}},\ }\bibfield  {title} {\bibinfo {title} {{Work output and thermal efficiency of an endoreversible entangled quantum Stirling engine with one dimensional isotropic Heisenberg model}},\ }\href {https://doi.org/https://doi.org/10.1016/j.physa.2019.123856} {\bibfield  {journal} {\bibinfo  {journal} {Physica A: Statistical Mechanics and its Applications}\ }\textbf {\bibinfo {volume} {547}},\ \bibinfo {pages} {123856} (\bibinfo {year} {2020})}\BibitemShut {NoStop}%
\bibitem [{\citenamefont {Zhang}(2008)}]{zhang2008entangled}%
  \BibitemOpen
  \bibfield  {author} {\bibinfo {author} {\bibfnamefont {G.~F.}\ \bibnamefont {Zhang}},\ }\bibfield  {title} {\bibinfo {title} {{Entangled quantum heat engines based on two two-spin systems with Dzyaloshinski-Moriya anisotropic antisymmetric interaction}},\ }\href {https://doi.org/10.1140/epjd/e2008-00133-0} {\bibfield  {journal} {\bibinfo  {journal} {The European Physical Journal D}\ }\textbf {\bibinfo {volume} {49}},\ \bibinfo {pages} {123} (\bibinfo {year} {2008})}\BibitemShut {NoStop}%
\bibitem [{\citenamefont {He}\ \emph {et~al.}(2012)\citenamefont {He}, \citenamefont {He},\ and\ \citenamefont {Zheng}}]{he2012thermal}%
  \BibitemOpen
  \bibfield  {author} {\bibinfo {author} {\bibfnamefont {X.}~\bibnamefont {He}}, \bibinfo {author} {\bibfnamefont {J.}~\bibnamefont {He}},\ and\ \bibinfo {author} {\bibfnamefont {J.}~\bibnamefont {Zheng}},\ }\bibfield  {title} {\bibinfo {title} {Thermal entangled quantum heat engine},\ }\href {https://doi.org/https://doi.org/10.1016/j.physa.2012.07.050} {\bibfield  {journal} {\bibinfo  {journal} {Physica A: Statistical Mechanics and its Applications}\ }\textbf {\bibinfo {volume} {391}},\ \bibinfo {pages} {6594} (\bibinfo {year} {2012})}\BibitemShut {NoStop}%
\bibitem [{\citenamefont {Hardal}\ and\ \citenamefont {M{\"u}stecapl{\i}o{\u{g}}lu}(2015)}]{hardal2015superradiant}%
  \BibitemOpen
  \bibfield  {author} {\bibinfo {author} {\bibfnamefont {A.~{\"U}.}\ \bibnamefont {Hardal}}\ and\ \bibinfo {author} {\bibfnamefont {{\"O}.~E.}\ \bibnamefont {M{\"u}stecapl{\i}o{\u{g}}lu}},\ }\bibfield  {title} {\bibinfo {title} {Superradiant quantum heat engine},\ }\href {https://doi.org/https://doi.org/10.1038/srep12953} {\bibfield  {journal} {\bibinfo  {journal} {Scientific reports}\ }\textbf {\bibinfo {volume} {5}},\ \bibinfo {pages} {12953} (\bibinfo {year} {2015})}\BibitemShut {NoStop}%
\bibitem [{\citenamefont {Ro\ss{}nagel}\ \emph {et~al.}(2014)\citenamefont {Ro\ss{}nagel}, \citenamefont {Abah}, \citenamefont {Schmidt-Kaler}, \citenamefont {Singer},\ and\ \citenamefont {Lutz}}]{PhysRevLett.112.030602}%
  \BibitemOpen
  \bibfield  {author} {\bibinfo {author} {\bibfnamefont {J.}~\bibnamefont {Ro\ss{}nagel}}, \bibinfo {author} {\bibfnamefont {O.}~\bibnamefont {Abah}}, \bibinfo {author} {\bibfnamefont {F.}~\bibnamefont {Schmidt-Kaler}}, \bibinfo {author} {\bibfnamefont {K.}~\bibnamefont {Singer}},\ and\ \bibinfo {author} {\bibfnamefont {E.}~\bibnamefont {Lutz}},\ }\bibfield  {title} {\bibinfo {title} {{Nanoscale Heat Engine Beyond the Carnot Limit}},\ }\href {https://doi.org/10.1103/PhysRevLett.112.030602} {\bibfield  {journal} {\bibinfo  {journal} {Phys. Rev. Lett.}\ }\textbf {\bibinfo {volume} {112}},\ \bibinfo {pages} {030602} (\bibinfo {year} {2014})}\BibitemShut {NoStop}%
\bibitem [{\citenamefont {Huang}\ \emph {et~al.}(2012)\citenamefont {Huang}, \citenamefont {Wang},\ and\ \citenamefont {Yi}}]{PhysRevE.86.051105}%
  \BibitemOpen
  \bibfield  {author} {\bibinfo {author} {\bibfnamefont {X.~L.}\ \bibnamefont {Huang}}, \bibinfo {author} {\bibfnamefont {T.}~\bibnamefont {Wang}},\ and\ \bibinfo {author} {\bibfnamefont {X.~X.}\ \bibnamefont {Yi}},\ }\bibfield  {title} {\bibinfo {title} {Effects of reservoir squeezing on quantum systems and work extraction},\ }\href {https://doi.org/10.1103/PhysRevE.86.051105} {\bibfield  {journal} {\bibinfo  {journal} {Phys. Rev. E}\ }\textbf {\bibinfo {volume} {86}},\ \bibinfo {pages} {051105} (\bibinfo {year} {2012})}\BibitemShut {NoStop}%
\bibitem [{\citenamefont {Arias}\ \emph {et~al.}(2018)\citenamefont {Arias}, \citenamefont {de~Oliveira},\ and\ \citenamefont {Sarandy}}]{UnruhOttoEngine}%
  \BibitemOpen
  \bibfield  {author} {\bibinfo {author} {\bibfnamefont {E.}~\bibnamefont {Arias}}, \bibinfo {author} {\bibfnamefont {T.~R.}\ \bibnamefont {de~Oliveira}},\ and\ \bibinfo {author} {\bibfnamefont {M.~S.}\ \bibnamefont {Sarandy}},\ }\bibfield  {title} {\bibinfo {title} {{The Unruh quantum Otto engine}},\ }\href {https://doi.org/https://doi.org/10.1007/JHEP02(2018)168} {\bibfield  {journal} {\bibinfo  {journal} {J. High Energy Phys.}\ }\textbf {\bibinfo {volume} {02}}\bibinfo  {number} { (2018)},\ \bibinfo {pages} {168}}\BibitemShut {NoStop}%
\bibitem [{\citenamefont {Gray}\ and\ \citenamefont {Mann}(2018)}]{Finn.UnruhOtto}%
  \BibitemOpen
\bibfield  {number} {  }\bibfield  {author} {\bibinfo {author} {\bibfnamefont {F.}~\bibnamefont {Gray}}\ and\ \bibinfo {author} {\bibfnamefont {R.~B.}\ \bibnamefont {Mann}},\ }\bibfield  {title} {\bibinfo {title} {{Scalar and fermionic Unruh Otto engines}},\ }\href {https://doi.org/https://doi.org/10.1007/JHEP11(2018)174} {\bibfield  {journal} {\bibinfo  {journal} {J. High Energy Phys.}\ }\textbf {\bibinfo {volume} {11}}\bibinfo  {number} { (2018)},\ \bibinfo {pages} {174}}\BibitemShut {NoStop}%
\bibitem [{\citenamefont {Gallock-Yoshimura}\ \emph {et~al.}(2023)\citenamefont {Gallock-Yoshimura}, \citenamefont {Thakur},\ and\ \citenamefont {Mann}}]{Gallock2023Otto}%
  \BibitemOpen
\bibfield  {number} {  }\bibfield  {author} {\bibinfo {author} {\bibfnamefont {K.}~\bibnamefont {Gallock-Yoshimura}}, \bibinfo {author} {\bibfnamefont {V.}~\bibnamefont {Thakur}},\ and\ \bibinfo {author} {\bibfnamefont {R.~B.}\ \bibnamefont {Mann}},\ }\bibfield  {title} {\bibinfo {title} {{Quantum Otto engine driven by quantum fields}},\ }\href {https://doi.org/10.3389/fphy.2023.1287860} {\bibfield  {journal} {\bibinfo  {journal} {Frontiers in Physics}\ }\textbf {\bibinfo {volume} {11}},\ \bibinfo {pages} {1287860} (\bibinfo {year} {2023})}\BibitemShut {NoStop}%
\bibitem [{\citenamefont {Xu}\ and\ \citenamefont {Yung}(2020)}]{Xu.UnruhOtto.degenerate}%
  \BibitemOpen
  \bibfield  {author} {\bibinfo {author} {\bibfnamefont {H.}~\bibnamefont {Xu}}\ and\ \bibinfo {author} {\bibfnamefont {M.-H.}\ \bibnamefont {Yung}},\ }\bibfield  {title} {\bibinfo {title} {{Unruh quantum Otto heat engine with level degeneracy}},\ }\href {https://doi.org/https://doi.org/10.1016/j.physletb.2020.135201} {\bibfield  {journal} {\bibinfo  {journal} {Physics Letters B}\ }\textbf {\bibinfo {volume} {801}},\ \bibinfo {pages} {135201} (\bibinfo {year} {2020})}\BibitemShut {NoStop}%
\bibitem [{\citenamefont {Kane}\ and\ \citenamefont {Majhi}(2021)}]{Kane.entangled.Unruh.Otto}%
  \BibitemOpen
  \bibfield  {author} {\bibinfo {author} {\bibfnamefont {G.~R.}\ \bibnamefont {Kane}}\ and\ \bibinfo {author} {\bibfnamefont {B.~R.}\ \bibnamefont {Majhi}},\ }\bibfield  {title} {\bibinfo {title} {{Entangled quantum Unruh Otto engine is more efficient}},\ }\href {https://doi.org/10.1103/PhysRevD.104.L041701} {\bibfield  {journal} {\bibinfo  {journal} {Phys. Rev. D}\ }\textbf {\bibinfo {volume} {104}},\ \bibinfo {pages} {L041701} (\bibinfo {year} {2021})}\BibitemShut {NoStop}%
\bibitem [{\citenamefont {Barman}\ and\ \citenamefont {Majhi}(2022)}]{Barman.entangled.UnruhOtto}%
  \BibitemOpen
  \bibfield  {author} {\bibinfo {author} {\bibfnamefont {D.}~\bibnamefont {Barman}}\ and\ \bibinfo {author} {\bibfnamefont {B.~R.}\ \bibnamefont {Majhi}},\ }\bibfield  {title} {\bibinfo {title} {{Constructing an entangled Unruh Otto engine and its efficiency}},\ }\href {https://doi.org/https://doi.org/10.1007/JHEP05(2022)046} {\bibfield  {journal} {\bibinfo  {journal} {J. High Energy Phys.}\ }\textbf {\bibinfo {volume} {05}}\bibinfo  {number} { (2022)},\ \bibinfo {pages} {046}}\BibitemShut {NoStop}%
\bibitem [{\citenamefont {Mukherjee}\ \emph {et~al.}(2022)\citenamefont {Mukherjee}, \citenamefont {Gangopadhyay},\ and\ \citenamefont {Majumdar}}]{Mukherjee.UnruhOtto.boundary}%
  \BibitemOpen
\bibfield  {number} {  }\bibfield  {author} {\bibinfo {author} {\bibfnamefont {A.}~\bibnamefont {Mukherjee}}, \bibinfo {author} {\bibfnamefont {S.}~\bibnamefont {Gangopadhyay}},\ and\ \bibinfo {author} {\bibfnamefont {A.~S.}\ \bibnamefont {Majumdar}},\ }\bibfield  {title} {\bibinfo {title} {{Unruh quantum Otto engine in the presence of a reflecting boundary}},\ }\href {https://doi.org/https://doi.org/10.1007/JHEP09(2022)105} {\bibfield  {journal} {\bibinfo  {journal} {J. High Energy Phys.}\ }\textbf {\bibinfo {volume} {09}}\bibinfo  {number} { (2022)},\ \bibinfo {pages} {105}}\BibitemShut {NoStop}%
\bibitem [{\citenamefont {Chattopadhyay}\ and\ \citenamefont {Paul}(2019)}]{Chattopadhyay_2019}%
  \BibitemOpen
\bibfield  {number} {  }\bibfield  {author} {\bibinfo {author} {\bibfnamefont {P.}~\bibnamefont {Chattopadhyay}}\ and\ \bibinfo {author} {\bibfnamefont {G.}~\bibnamefont {Paul}},\ }\bibfield  {title} {\bibinfo {title} {Relativistic quantum heat engine from uncertainty relation standpoint},\ }\href {https://doi.org/10.1038/s41598-019-53331-x} {\bibfield  {journal} {\bibinfo  {journal} {Scientific Reports}\ }\textbf {\bibinfo {volume} {9}},\ \bibinfo {pages} {16967} (\bibinfo {year} {2019})}\BibitemShut {NoStop}%
\bibitem [{\citenamefont {Mu\~noz}\ and\ \citenamefont {Pe\~na}(2012)}]{PhysRevE.86.061108}%
  \BibitemOpen
  \bibfield  {author} {\bibinfo {author} {\bibfnamefont {E.}~\bibnamefont {Mu\~noz}}\ and\ \bibinfo {author} {\bibfnamefont {F.~J.}\ \bibnamefont {Pe\~na}},\ }\bibfield  {title} {\bibinfo {title} {Quantum heat engine in the relativistic limit: The case of a dirac particle},\ }\href {https://doi.org/10.1103/PhysRevE.86.061108} {\bibfield  {journal} {\bibinfo  {journal} {Phys. Rev. E}\ }\textbf {\bibinfo {volume} {86}},\ \bibinfo {pages} {061108} (\bibinfo {year} {2012})}\BibitemShut {NoStop}%
\bibitem [{\citenamefont {Sukamto}\ \emph {et~al.}(2023)\citenamefont {Sukamto}, \citenamefont {Yuwana},\ and\ \citenamefont {Purwanto}}]{Sukamto_2023}%
  \BibitemOpen
  \bibfield  {author} {\bibinfo {author} {\bibfnamefont {H.}~\bibnamefont {Sukamto}}, \bibinfo {author} {\bibfnamefont {L.}~\bibnamefont {Yuwana}},\ and\ \bibinfo {author} {\bibfnamefont {A.}~\bibnamefont {Purwanto}},\ }\bibfield  {title} {\bibinfo {title} {Relativistic quantum heat engine with the presence of minimal length},\ }\href {https://doi.org/10.1088/1402-4896/acec1f} {\bibfield  {journal} {\bibinfo  {journal} {Physica Scripta}\ }\textbf {\bibinfo {volume} {98}},\ \bibinfo {pages} {095403} (\bibinfo {year} {2023})}\BibitemShut {NoStop}%
\bibitem [{\citenamefont {Myers}\ \emph {et~al.}(2021)\citenamefont {Myers}, \citenamefont {Abah},\ and\ \citenamefont {Deffner}}]{Myers_2021}%
  \BibitemOpen
  \bibfield  {author} {\bibinfo {author} {\bibfnamefont {N.~M.}\ \bibnamefont {Myers}}, \bibinfo {author} {\bibfnamefont {O.}~\bibnamefont {Abah}},\ and\ \bibinfo {author} {\bibfnamefont {S.}~\bibnamefont {Deffner}},\ }\bibfield  {title} {\bibinfo {title} {Quantum otto engines at relativistic energies},\ }\href {https://doi.org/10.1088/1367-2630/ac2756} {\bibfield  {journal} {\bibinfo  {journal} {New Journal of Physics}\ }\textbf {\bibinfo {volume} {23}},\ \bibinfo {pages} {105001} (\bibinfo {year} {2021})}\BibitemShut {NoStop}%
\bibitem [{\citenamefont {Purwanto}\ \emph {et~al.}(2016)\citenamefont {Purwanto}, \citenamefont {Sukamto}, \citenamefont {Subagyo},\ and\ \citenamefont {Taufiqi}}]{Purwanto_2016}%
  \BibitemOpen
  \bibfield  {author} {\bibinfo {author} {\bibfnamefont {A.}~\bibnamefont {Purwanto}}, \bibinfo {author} {\bibfnamefont {H.}~\bibnamefont {Sukamto}}, \bibinfo {author} {\bibfnamefont {B.~A.}\ \bibnamefont {Subagyo}},\ and\ \bibinfo {author} {\bibfnamefont {M.}~\bibnamefont {Taufiqi}},\ }\bibfield  {title} {\bibinfo {title} {Two scenarios on the relativistic quantum heat engine},\ }\bibfield  {journal} {\bibinfo  {journal} {Journal of Applied Mathematics and Physics}\ }\textbf {\bibinfo {volume} {5}},\ \href {https://doi.org/10.4236/jamp.2016.47144} {10.4236/jamp.2016.47144} (\bibinfo {year} {2016})\BibitemShut {NoStop}%
\bibitem [{\citenamefont {Pe\~na}\ \emph {et~al.}(2016)\citenamefont {Pe\~na}, \citenamefont {Ferr\'e}, \citenamefont {Orellana}, \citenamefont {Rojas},\ and\ \citenamefont {Vargas}}]{PhysRevE.94.022109}%
  \BibitemOpen
  \bibfield  {author} {\bibinfo {author} {\bibfnamefont {F.~J.}\ \bibnamefont {Pe\~na}}, \bibinfo {author} {\bibfnamefont {M.}~\bibnamefont {Ferr\'e}}, \bibinfo {author} {\bibfnamefont {P.~A.}\ \bibnamefont {Orellana}}, \bibinfo {author} {\bibfnamefont {R.~G.}\ \bibnamefont {Rojas}},\ and\ \bibinfo {author} {\bibfnamefont {P.}~\bibnamefont {Vargas}},\ }\bibfield  {title} {\bibinfo {title} {Optimization of a relativistic quantum mechanical engine},\ }\href {https://doi.org/10.1103/PhysRevE.94.022109} {\bibfield  {journal} {\bibinfo  {journal} {Phys. Rev. E}\ }\textbf {\bibinfo {volume} {94}},\ \bibinfo {pages} {022109} (\bibinfo {year} {2016})}\BibitemShut {NoStop}%
\bibitem [{\citenamefont {Gallock-Yoshimura}(2024)}]{Gallock.Otto.delta}%
  \BibitemOpen
  \bibfield  {author} {\bibinfo {author} {\bibfnamefont {K.}~\bibnamefont {Gallock-Yoshimura}},\ }\bibfield  {title} {\bibinfo {title} {{Relativistic quantum Otto engine: instant work extraction from a quantum field}},\ }\href {https://doi.org/https://doi.org/10.1007/JHEP01(2024)198} {\bibfield  {journal} {\bibinfo  {journal} {J. High Energy Phys.}\ }\textbf {\bibinfo {volume} {01}}\bibinfo  {number} { (2024)},\ \bibinfo {pages} {1998}}\BibitemShut {NoStop}%
\bibitem [{\citenamefont {Kollas}\ and\ \citenamefont {Moustos}(2024)}]{Kollas.Otto.2024}%
  \BibitemOpen
\bibfield  {number} {  }\bibfield  {author} {\bibinfo {author} {\bibfnamefont {N.~K.}\ \bibnamefont {Kollas}}\ and\ \bibinfo {author} {\bibfnamefont {D.}~\bibnamefont {Moustos}},\ }\bibfield  {title} {\bibinfo {title} {{An exactly solvable relativistic quantum Otto engine}},\ }\href {https://doi.org/10.1103/PhysRevD.109.065025} {\bibfield  {journal} {\bibinfo  {journal} {Phys. Rev. D}\ }\textbf {\bibinfo {volume} {109}},\ \bibinfo {pages} {065025} (\bibinfo {year} {2024})}\BibitemShut {NoStop}%
\bibitem [{\citenamefont {Unruh}(1976)}]{Unruh1979evaporation}%
  \BibitemOpen
  \bibfield  {author} {\bibinfo {author} {\bibfnamefont {W.~G.}\ \bibnamefont {Unruh}},\ }\bibfield  {title} {\bibinfo {title} {Notes on black-hole evaporation},\ }\href {https://doi.org/10.1103/PhysRevD.14.870} {\bibfield  {journal} {\bibinfo  {journal} {Phys. Rev. D}\ }\textbf {\bibinfo {volume} {14}},\ \bibinfo {pages} {870} (\bibinfo {year} {1976})}\BibitemShut {NoStop}%
\bibitem [{\citenamefont {{DeWitt}}(1979)}]{DeWitt1979}%
  \BibitemOpen
  \bibfield  {author} {\bibinfo {author} {\bibfnamefont {B.~S.}\ \bibnamefont {{DeWitt}}},\ }\bibfield  {title} {\bibinfo {title} {{Quantum gravity: The new synthesis}},\ }in\ \href@noop {} {\emph {\bibinfo {booktitle} {General Relativity: An Einstein Centenary Survey}}},\ \bibinfo {editor} {edited by\ \bibinfo {editor} {\bibfnamefont {S.~W.}\ \bibnamefont {{Hawking}}}\ and\ \bibinfo {editor} {\bibfnamefont {W.}~\bibnamefont {{Israel}}}}\ (\bibinfo  {publisher} {Cambridge University Press},\ \bibinfo {year} {1979})\ pp.\ \bibinfo {pages} {680--745}\BibitemShut {NoStop}%
\bibitem [{\citenamefont {Mart\'{\i}n-Mart\'{\i}nez}\ \emph {et~al.}(2013)\citenamefont {Mart\'{\i}n-Mart\'{\i}nez}, \citenamefont {Montero},\ and\ \citenamefont {del Rey}}]{PhysRevD.87.064038}%
  \BibitemOpen
  \bibfield  {author} {\bibinfo {author} {\bibfnamefont {E.}~\bibnamefont {Mart\'{\i}n-Mart\'{\i}nez}}, \bibinfo {author} {\bibfnamefont {M.}~\bibnamefont {Montero}},\ and\ \bibinfo {author} {\bibfnamefont {M.}~\bibnamefont {del Rey}},\ }\bibfield  {title} {\bibinfo {title} {Wavepacket detection with the unruh-dewitt model},\ }\href {https://doi.org/10.1103/PhysRevD.87.064038} {\bibfield  {journal} {\bibinfo  {journal} {Phys. Rev. D}\ }\textbf {\bibinfo {volume} {87}},\ \bibinfo {pages} {064038} (\bibinfo {year} {2013})}\BibitemShut {NoStop}%
\bibitem [{\citenamefont {Funai}\ \emph {et~al.}(2019)\citenamefont {Funai}, \citenamefont {Louko},\ and\ \citenamefont {Mart\'{\i}n-Mart\'{\i}nez}}]{Funai.lightmatter.2019}%
  \BibitemOpen
  \bibfield  {author} {\bibinfo {author} {\bibfnamefont {N.}~\bibnamefont {Funai}}, \bibinfo {author} {\bibfnamefont {J.}~\bibnamefont {Louko}},\ and\ \bibinfo {author} {\bibfnamefont {E.}~\bibnamefont {Mart\'{\i}n-Mart\'{\i}nez}},\ }\bibfield  {title} {\bibinfo {title} {{$\hat{\bm p}\cdot\hat{\bm{A}}$} vs {$\hat{\bm x}\cdot\hat{\bm{E}}$}: Gauge invariance in quantum optics and quantum field theory},\ }\href {https://doi.org/10.1103/PhysRevD.99.065014} {\bibfield  {journal} {\bibinfo  {journal} {Phys. Rev. D}\ }\textbf {\bibinfo {volume} {99}},\ \bibinfo {pages} {065014} (\bibinfo {year} {2019})}\BibitemShut {NoStop}%
\bibitem [{\citenamefont {Lopp}\ and\ \citenamefont {Mart\'{\i}n-Mart\'{\i}nez}(2021)}]{Lopp.lightmatter.2021}%
  \BibitemOpen
  \bibfield  {author} {\bibinfo {author} {\bibfnamefont {R.}~\bibnamefont {Lopp}}\ and\ \bibinfo {author} {\bibfnamefont {E.}~\bibnamefont {Mart\'{\i}n-Mart\'{\i}nez}},\ }\bibfield  {title} {\bibinfo {title} {Quantum delocalization, gauge, and quantum optics: Light-matter interaction in relativistic quantum information},\ }\href {https://doi.org/10.1103/PhysRevA.103.013703} {\bibfield  {journal} {\bibinfo  {journal} {Phys. Rev. A}\ }\textbf {\bibinfo {volume} {103}},\ \bibinfo {pages} {013703} (\bibinfo {year} {2021})}\BibitemShut {NoStop}%
\bibitem [{\citenamefont {Shah}\ \emph {et~al.}(2025)\citenamefont {Shah}, \citenamefont {Mart\'{\i}n-Mart\'{\i}nez},\ and\ \citenamefont {Perche}}]{Shah.UDW.spin.2025}%
  \BibitemOpen
  \bibfield  {author} {\bibinfo {author} {\bibfnamefont {R.}~\bibnamefont {Shah}}, \bibinfo {author} {\bibfnamefont {E.}~\bibnamefont {Mart\'{\i}n-Mart\'{\i}nez}},\ and\ \bibinfo {author} {\bibfnamefont {T.~R.}\ \bibnamefont {Perche}},\ }\bibfield  {title} {\bibinfo {title} {{Relativistic QFT description for the interaction of a spin with a magnetic field}},\ }\href {https://doi.org/10.1103/PhysRevD.111.044075} {\bibfield  {journal} {\bibinfo  {journal} {Phys. Rev. D}\ }\textbf {\bibinfo {volume} {111}},\ \bibinfo {pages} {044075} (\bibinfo {year} {2025})}\BibitemShut {NoStop}%
\bibitem [{\citenamefont {You}\ and\ \citenamefont {Nori}(2011)}]{You:2011}%
  \BibitemOpen
  \bibfield  {author} {\bibinfo {author} {\bibfnamefont {J.~Q.}\ \bibnamefont {You}}\ and\ \bibinfo {author} {\bibfnamefont {F.}~\bibnamefont {Nori}},\ }\bibfield  {title} {\bibinfo {title} {{Atomic physics and quantum optics using superconducting circuits}},\ }\href {https://doi.org/10.1038/nature10122} {\bibfield  {journal} {\bibinfo  {journal} {Nature}\ }\textbf {\bibinfo {volume} {474}},\ \bibinfo {pages} {589} (\bibinfo {year} {2011})}\BibitemShut {NoStop}%
\bibitem [{\citenamefont {Lima}\ \emph {et~al.}(2023)\citenamefont {Lima}, \citenamefont {Patterson}, \citenamefont {Tjoa},\ and\ \citenamefont {Mann}}]{Lima.Unruh.qutrit}%
  \BibitemOpen
  \bibfield  {author} {\bibinfo {author} {\bibfnamefont {C.}~\bibnamefont {Lima}}, \bibinfo {author} {\bibfnamefont {E.}~\bibnamefont {Patterson}}, \bibinfo {author} {\bibfnamefont {E.}~\bibnamefont {Tjoa}},\ and\ \bibinfo {author} {\bibfnamefont {R.~B.}\ \bibnamefont {Mann}},\ }\bibfield  {title} {\bibinfo {title} {Unruh phenomena and thermalization for qudit detectors},\ }\href {https://doi.org/10.1103/PhysRevD.108.105020} {\bibfield  {journal} {\bibinfo  {journal} {Phys. Rev. D}\ }\textbf {\bibinfo {volume} {108}},\ \bibinfo {pages} {105020} (\bibinfo {year} {2023})}\BibitemShut {NoStop}%
\bibitem [{\citenamefont {Gallock-Yoshimura}\ \emph {et~al.}(2025)\citenamefont {Gallock-Yoshimura}, \citenamefont {Osawa},\ and\ \citenamefont {Nambu}}]{Ken.gapless.qudit.2025}%
  \BibitemOpen
  \bibfield  {author} {\bibinfo {author} {\bibfnamefont {K.}~\bibnamefont {Gallock-Yoshimura}}, \bibinfo {author} {\bibfnamefont {Y.}~\bibnamefont {Osawa}},\ and\ \bibinfo {author} {\bibfnamefont {Y.}~\bibnamefont {Nambu}},\ }\bibfield  {title} {\bibinfo {title} {{Acceleration-induced radiation from a qudit particle detector model}},\ }\href {https://doi.org/10.1103/PhysRevD.111.105012} {\bibfield  {journal} {\bibinfo  {journal} {Phys. Rev. D}\ }\textbf {\bibinfo {volume} {111}},\ \bibinfo {pages} {105012} (\bibinfo {year} {2025})}\BibitemShut {NoStop}%
\bibitem [{\citenamefont {Sonkar}\ and\ \citenamefont {Johal}(2024)}]{sonkar2024energygap}%
  \BibitemOpen
  \bibfield  {author} {\bibinfo {author} {\bibfnamefont {S.}~\bibnamefont {Sonkar}}\ and\ \bibinfo {author} {\bibfnamefont {R.~S.}\ \bibnamefont {Johal}},\ }\bibfield  {title} {\bibinfo {title} {{Energy-gap modulation and majorization in three-level quantum Otto engine}},\ }\href {https://arxiv.org/abs/2403.09154} {\bibfield  {journal} {\bibinfo  {journal} {arXiv:2403.09154}\ } (\bibinfo {year} {2024})}\BibitemShut {NoStop}%
\bibitem [{\citenamefont {Anka}\ \emph {et~al.}(2021)\citenamefont {Anka}, \citenamefont {de~Oliveira},\ and\ \citenamefont {Jonathan}}]{Anka.multilevel.2021}%
  \BibitemOpen
  \bibfield  {author} {\bibinfo {author} {\bibfnamefont {M.~F.}\ \bibnamefont {Anka}}, \bibinfo {author} {\bibfnamefont {T.~R.}\ \bibnamefont {de~Oliveira}},\ and\ \bibinfo {author} {\bibfnamefont {D.}~\bibnamefont {Jonathan}},\ }\bibfield  {title} {\bibinfo {title} {{Measurement-based quantum heat engine in a multilevel system}},\ }\href {https://doi.org/10.1103/PhysRevE.104.054128} {\bibfield  {journal} {\bibinfo  {journal} {Phys. Rev. E}\ }\textbf {\bibinfo {volume} {104}},\ \bibinfo {pages} {054128} (\bibinfo {year} {2021})}\BibitemShut {NoStop}%
\bibitem [{\citenamefont {Griffiths}(2005)}]{Griffiths2005QM.2nd}%
  \BibitemOpen
  \bibfield  {author} {\bibinfo {author} {\bibfnamefont {D.~J.}\ \bibnamefont {Griffiths}},\ }\href@noop {} {\emph {\bibinfo {title} {Introduction to Quantum Mechanics. 2nd ed.}}}\ (\bibinfo  {publisher} {Pearson, Upper Saddle River, New Jersey, USA},\ \bibinfo {year} {2005})\BibitemShut {NoStop}%
\bibitem [{\citenamefont {Alicki}(1979)}]{R.Alicki_1979}%
  \BibitemOpen
  \bibfield  {author} {\bibinfo {author} {\bibfnamefont {R.}~\bibnamefont {Alicki}},\ }\bibfield  {title} {\bibinfo {title} {The quantum open system as a model of the heat engine},\ }\href {https://doi.org/10.1088/0305-4470/12/5/007} {\bibfield  {journal} {\bibinfo  {journal} {Journal of Physics A: Mathematical and General}\ }\textbf {\bibinfo {volume} {12}},\ \bibinfo {pages} {L103} (\bibinfo {year} {1979})}\BibitemShut {NoStop}%
\bibitem [{\citenamefont {Fewster}\ \emph {et~al.}(2016)\citenamefont {Fewster}, \citenamefont {Juárez-Aubry},\ and\ \citenamefont {Louko}}]{Fewster.Waiting.Unruh}%
  \BibitemOpen
  \bibfield  {author} {\bibinfo {author} {\bibfnamefont {C.~J.}\ \bibnamefont {Fewster}}, \bibinfo {author} {\bibfnamefont {B.~A.}\ \bibnamefont {Juárez-Aubry}},\ and\ \bibinfo {author} {\bibfnamefont {J.}~\bibnamefont {Louko}},\ }\bibfield  {title} {\bibinfo {title} {{Waiting for Unruh}},\ }\href {https://doi.org/https://doi.org/10.1088/0264-9381/33/16/165003} {\bibfield  {journal} {\bibinfo  {journal} {Class. Quantum Grav.}\ }\textbf {\bibinfo {volume} {33}},\ \bibinfo {pages} {165003} (\bibinfo {year} {2016})}\BibitemShut {NoStop}%
\bibitem [{\citenamefont {Garay}\ \emph {et~al.}(2016)\citenamefont {Garay}, \citenamefont {Mart\'{\i}n-Mart\'{\i}nez},\ and\ \citenamefont {de~Ram\'on}}]{Garay2016anti-unruh}%
  \BibitemOpen
  \bibfield  {author} {\bibinfo {author} {\bibfnamefont {L.~J.}\ \bibnamefont {Garay}}, \bibinfo {author} {\bibfnamefont {E.}~\bibnamefont {Mart\'{\i}n-Mart\'{\i}nez}},\ and\ \bibinfo {author} {\bibfnamefont {J.}~\bibnamefont {de~Ram\'on}},\ }\bibfield  {title} {\bibinfo {title} {{Thermalization of particle detectors: The Unruh effect and its reverse}},\ }\href {https://doi.org/10.1103/PhysRevD.94.104048} {\bibfield  {journal} {\bibinfo  {journal} {Phys. Rev. D}\ }\textbf {\bibinfo {volume} {94}},\ \bibinfo {pages} {104048} (\bibinfo {year} {2016})}\BibitemShut {NoStop}%
\bibitem [{\citenamefont {Shi}\ \emph {et~al.}(2020)\citenamefont {Shi}, \citenamefont {Shi}, \citenamefont {Wang}, \citenamefont {Hu}, \citenamefont {Liu}, \citenamefont {Yang},\ and\ \citenamefont {Fan}}]{shi2020quantum.coherence}%
  \BibitemOpen
  \bibfield  {author} {\bibinfo {author} {\bibfnamefont {Y.-H.}\ \bibnamefont {Shi}}, \bibinfo {author} {\bibfnamefont {H.-L.}\ \bibnamefont {Shi}}, \bibinfo {author} {\bibfnamefont {X.-H.}\ \bibnamefont {Wang}}, \bibinfo {author} {\bibfnamefont {M.-L.}\ \bibnamefont {Hu}}, \bibinfo {author} {\bibfnamefont {S.-Y.}\ \bibnamefont {Liu}}, \bibinfo {author} {\bibfnamefont {W.-L.}\ \bibnamefont {Yang}},\ and\ \bibinfo {author} {\bibfnamefont {H.}~\bibnamefont {Fan}},\ }\bibfield  {title} {\bibinfo {title} {Quantum coherence in a quantum heat engine},\ }\href {https://doi.org/10.1088/1751-8121/ab6a6b} {\bibfield  {journal} {\bibinfo  {journal} {Journal of Physics A: Mathematical and Theoretical}\ }\textbf {\bibinfo {volume} {53}},\ \bibinfo {pages} {085301} (\bibinfo {year} {2020})}\BibitemShut {NoStop}%
\bibitem [{\citenamefont {Bernardo}(2020)}]{B.Lima}%
  \BibitemOpen
  \bibfield  {author} {\bibinfo {author} {\bibfnamefont {B.~d.~L.}\ \bibnamefont {Bernardo}},\ }\bibfield  {title} {\bibinfo {title} {Unraveling the role of coherence in the first law of quantum thermodynamics},\ }\href {https://doi.org/10.1103/PhysRevE.102.062152} {\bibfield  {journal} {\bibinfo  {journal} {Phys. Rev. E}\ }\textbf {\bibinfo {volume} {102}},\ \bibinfo {pages} {062152} (\bibinfo {year} {2020})}\BibitemShut {NoStop}%
\end{thebibliography}%

\end{document}